\documentclass[lettersize,journal]{IEEEtran}
\usepackage{amsmath,amsfonts}
\usepackage{array}
\usepackage{textcomp}
\usepackage{stfloats}
\usepackage{float}
\usepackage{url}
\usepackage{verbatim}
\usepackage{graphicx}
\usepackage{cite}
\usepackage{xspace}
\usepackage{hyperref}

\usepackage{wrapfig}
\usepackage{amsthm}
\usepackage{multirow, multicol}
\usepackage{enumitem,kantlipsum}

\usepackage{arydshln}
\newcommand{\num}[1]{\relax\ifmmode \mathbb #1\else $\mathbb #1$\fi}
\newcommand{\reals}{{\num R}}

\newcommand{\deq}{\doteq}

\usepackage{algorithmic}
\usepackage[linesnumbered,lined,commentsnumbered,norelsize]{algorithm2e}
\newtheorem{definition}{\textbf{Definition}}
\newtheorem{theorem}{\textbf{Theorem}}
\newtheorem{lemma}{\textbf{Lemma}}

\newtheorem{example}{Example}
\newtheorem{remark}{Remark}
\usepackage{subfigure}

\SetCommentSty{mycommfont}
\usepackage{xcolor}

\usepackage{cite}
\usepackage{amsmath,amssymb,amsfonts}

\usepackage{algorithmic}

\usepackage{float}
\usepackage{wrapfig}
\usepackage{graphicx}
\usepackage{subfigure}
\RequirePackage{amsmath, dsfont, amssymb}
\usepackage{bm}
\newcommand{\Star}{S}
\newcommand{\nnnum}[1]{\relax\ifmmode 
  {\mathbb #1}_{\geq 0} \else ${\mathbb #1}_{\geq 0}$
  \fi}
\newcommand{\state}[1]{#1}
\newcommand{\scalar}[1]{\mathbf{#1}}
\newcommand{\unsafe}{\Psi}
\newcommand{\timebound}{{\mathrm{T}}}
\newcommand{\reachset}{{Reach(\mathcal{H}, \Theta,\mathcal{U},\timebound)}}
\newcommand{\algfnt}[1]{\textrm{#1}}
\newcommand{\algnm}[1]{\textcolor{blue}{#1}}

\newcommand{\tup}[1]
           {
             \relax\ifmmode
             \langle #1 \rangle
             \else $\langle$ #1 $\rangle$ \fi
           }
\newcommand{\meansl}{[\![}
\newcommand{\meansr}{]\!]}
\newcommand{\means}[1]{\meansl #1 \meansr}

\graphicspath{Figures/}
\usepackage[figurename=Figure]{caption}
\usepackage{multirow}
\usepackage{arydshln}
\usepackage{hyperref}

\def\BibTeX{{\rm B\kern-.05em{\sc i\kern-.025em b}\kern-.08em
    T\kern-.1667em\lower.7ex\hbox{E}\kern-.125emX}}
\begin{document}
\title{BDD for Complete Characterization of a Safety Violation in Linear Systems with Inputs}
\author{Manish Goyal, David Bergman, and Parasara Sridhar Duggirala
\thanks{M. Goyal is with the Computer Science Department at University of North Carolina at Chapel Hill, NC 27516 USA. (e-mail: manishg@cs.unc.edu). }
\thanks{D. Bergman is with the School of Business at University of Connecticut, Storrs CT 06229 USA. (e-mail: david.bergman@uconn.edu).}
\thanks{P. S. Duggirala is with the Computer Science Department at University of North Carolina at Chapel Hill, NC 27516 USA. (e-mail: psd@cs.unc.edu).}}

\maketitle
\begin{abstract}
The control design tools for linear systems typically involves pole placement and computing Lyapunov functions which are useful for ensuring stability. But given higher requirements on control design, a designer is expected to satisfy other specification such as safety or temporal logic specification as well, and a naive control design might not satisfy such specification.
A control designer can employ model checking as a tool for checking safety and obtain a counterexample in case of a safety violation. 
While several scalable techniques for verification have been developed for safety verification of linear dynamical systems, such tools merely act as decision procedures to evaluate system safety and, consequently, yield a counterexample as an evidence to safety violation. However these model checking methods are not geared towards discovering corner cases or re-using verification artifacts for another sub-optimal safety specification.
%
%
In this paper, we describe a technique for obtaining complete characterization of counterexamples for a safety violation in linear systems. The proposed technique uses the reachable set computed during safety verification for a given temporal logic formula, performs constraint propagation, and represents all modalities of counterexamples using a binary decision diagram (BDD). We introduce an approach to dynamically determine isomorphic nodes for obtaining a considerably reduced (in size) decision diagram. A thorough experimental evaluation on various benchmarks exhibits that the reduction technique achieves up to $67\%$ reduction in the number of nodes and $75\%$ reduction in the width of the decision diagram.
\end{abstract}

\textit{Index terms-} Safety verification, linear system, binary decision diagram, optimization, mixed integer linear program
\section{Introduction}

Integration of software with embedded control systems has evolved as a field of Cyber-physical systems (CPS). It involves interaction between continuous physical environment modeled as an ordinary differential equation (ODE) and discrete software systems, which requires sensing and controlling physical quantities.
Designing a controller for these (potentially infinite) state systems is  an iterative process so that the system meets the desired behaviors such as stability, robustness, or at best safety~\cite{DBLP:journals/corr/abs-2107-05815}. 
For a given system model and specification, the control designer uses tools in their repertoire to come up with a controller, checks if the system satisfies the required specification and iteratively refines the controller. Stability and safety are two important classes of specifications. One has to come up with either a common~\cite{narendra1994common,liberzon1999basic,lin2009stability} or multiple ~\cite{branicky1998multiple,tanaka2003multiple} Lyapunov function(s) to prove stability, whereas a model checking tool~\cite{spaceex2011,bak2017hylaa,goyal2012reachability,goyal2011translationVerimag} is employed to verify a system with respect to a safety specification. A safety specification is satisfied if all executions of a system avoid the set of states labelled as \emph{unsafe}.
Any execution that encounters an unsafe state is called a \emph{counterexample} to the safety specification.

While the tools for stability provide intuitive information to the designer, 
most of the model checking approaches for safety verification focus on computing over-approximation of reachable set and hence establish the safety specification. They typically yield one counterexample as an evidence to safety violation. However, one counterexample may not give much insight to a control designer when in practice they are more interested in identifying corner cases of a safety violation. 
In contrast to the work-flow where a designer come up with a stable control and then checks its safety, supplementing the analysis with a complete characterization of counterexamples can be helpful for control designer for various reasons. First, even a small change in a safety specification due to operating condition can render an originally safe controller unsafe. But it may not be advisable to conduct verification  afresh because performing safety analysis is time consuming. 
Second, not all counterexamples to the safety specification are equivalent to a control designer. 
Finally, multiple characterizations can further aid the design process by providing the intuition behind a safety violation.


Originally, counterexamples were a mere side effect of model checking. More recently, techniques were developed to uncover deep bugs which would otherwise take a long time to uncover~\cite{bradley2011sat,bradley2012ic3,DBLP:conf/atva/0002D20,nexg-emsoft-2022,DBLP:phd/basesearch/Goyal22,goyalpsd_dars_2019,goyal_l4dc_2020}. 
The introduction of \emph{Counter-Example-Guided-Abstraction-Refinement} (CEGAR)~\cite{cgjlv00} changed the role of counterexamples from a mere feature to an algorithmic tool, where the counterexample acts as a primary guide to restrict the space of possible refinements. In the domain of automated synthesis, \emph{Counterexample Guided Inductive Synthesis} (CEGIS)  framework~\cite{Solar-LezamaTBSS06,solar2008program}, as the name suggests, leverages counterexamples from verification for inductive synthesis. The characterization of counterexamples can potentially aid these abstraction refinement and synthesis frameworks by providing them with multiple counterexamples.

%

\begin{figure}[ht]
\centering
\includegraphics[width=0.45\textwidth]{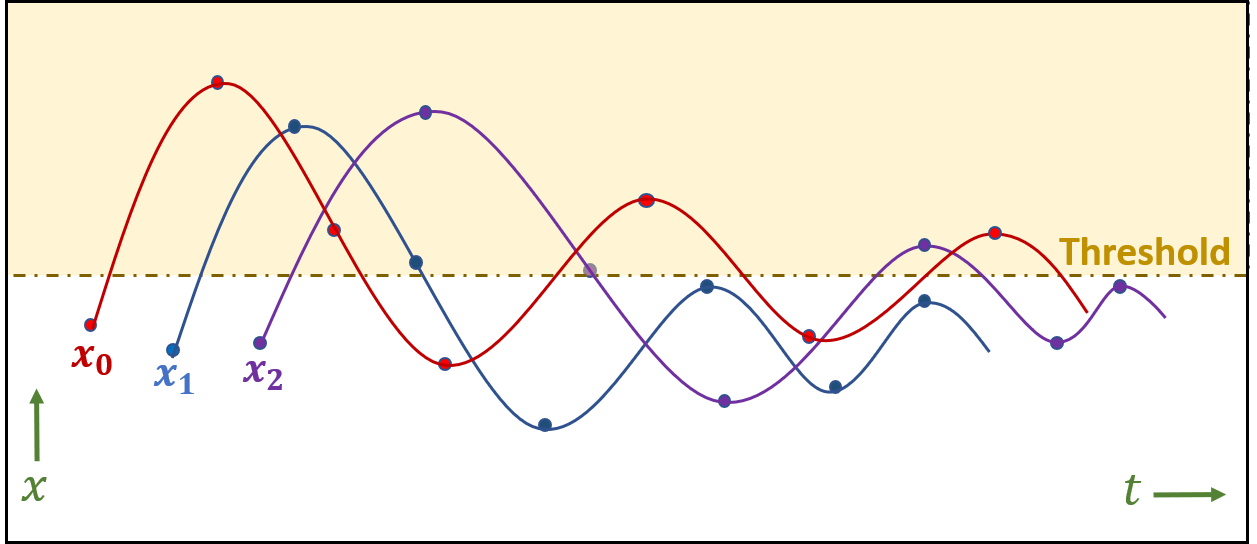}
\caption{\small{Different profiles of the system overshooting the threshold. The designer might be interested in obtaining those executions which cross the threshold at multiple (possibly non-contiguous) time steps, at just one time instance, or at maximum number of time steps etc.}}
\label{fig:intro-reg-control}
\end{figure}

Metrics over counterexamples can be used as a proxy for comparing performance of controllers across multiple iterations of control design and analysis. 
Such metrics can also be useful when the cost function for designing optimal control is non-convex.
A few approaches have been developed for generating different, namely, \emph{longest}, \emph{deepest}, and \emph{robust} counterexamples in linear hybrid systems~\cite{DBLP:conf/adhs/GoyalD18,DBLP:journals/automatica/GoyalD20,DBLP:conf/amcc/0002BD20}. But these counterexamples fail to  capture all modalities of a safety violation. Consider  multiple executions of a system in Fig.~\ref{fig:intro-reg-control}. As shown, all executions cross the threshold where this overlap with the threshold could denote an overshoot in the regulation control, an undesirable maneuver in an autonomous car, a failure site in hardware design, or similar specification violation in a given system. The dots indicate executions states at certain discrete time steps. Although it is important to observe that the system can go above the threshold, the control designer might be  interested in executions crossing it multiple (possibly non-contiguous) time steps or more precisely how many times was it crossed. Currently, none of the existing safety verification tools is equipped with such functionality.
%
%

Our motivation to achieve complete characterization of a safety violation in control systems is inspired by failure identification in computer aided design, where extracting the essence of an error may still require a great deal of human effort. Yet debugging a design is shown to be greatly benefited from using more than one counterexample as well as by efficiently identifying crucial sites leading to the failure. We aim to obtain complete characterization of counterexamples for linear dynamical  systems based on these insights.

The presented work builds on the prior work of computing a reachable set~\cite{bak2017rigorous}, which includes the set of states encountered by a simulation algorithm for systems with linear hybrid dynamics.
The reachable set computation leverages the superposition principle and the generalized star representation~\cite{bak2017rigorous, duggirala2016parsimonious}. Our counterexample generation mechanism reuses the artifacts generated during the model checking process and employs constraint propagation. Then it constructs a graphical structure called \emph{binary decision diagram} (BDD) for representing all modalities of a safety violation.

The main contributions of this paper is an efficient technique for obtaining complete characterization of counterexamples to a temporal logic safety specification in the domain of linear (possibly infinite state) systems with piece-wise bounded inputs. To be more precise - (i) it first proposes a framework for constructing this characterization in the form of a BDD, (ii) then, it introduces an approach for examining node isomorphism leading to a much more succinct BDD representation, (iii) it provides a Mixed Integer Linear Program (MILP) formulation of the system that models node isomorphism,  and (iv) finally, it performs through evaluation of presented   techniques on numerous benchmarks.

%
%
\section{Related Work}
\label{sec:relwork}
Generating specific type of counterexamples has been an active research topic in model checking. In some recent works~\cite{DBLP:conf/adhs/GoyalD18,DBLP:journals/automatica/GoyalD20,DBLP:conf/amcc/0002BD20,DBLP:phd/basesearch/Goyal22}, the authors provide techniques to generate a variety of counterexamples for linear dynamical systems. In the domain of hybrid systems, many CEGAR based approaches pursue various notions of counterexamples~\cite{fcjk05,dkl07,alur2003counter,prabhakar2013hybrid,DM:hystab:ICCPS2011,ratschan2005safety,sankaranarayanan2011relational}. Most of them are restricted to the domain of timed and rectangular hybrid systems. The current state-of-the-art tools such as SpaceEx~\cite{spaceex1,spaceex2011} and HyLAA~\cite{bak2017hylaa} spit out the counterexample that violates the safety specification at the earliest time and at the latest time, respectively.

Counterexamples also play an important role in {\em falsification} techniques~\cite{Fainekos09,DonzeM10}. Instead of proving that the specification is satisfied, falsification tools like S-Taliro~\cite{AnnpureddyLFS11} and Breach~\cite{BreachAD} search for an execution that violates the specification. Given a safety specification of cyber-physical system in Metric Temporal Logic (MTL)~\cite{MTLmain} or Signal Temporal Logic (STL)~\cite{STLmain}, falsification tools employ a variety of techniques~\cite{MCFalsification,AbbasF11,zutshi2014multiple,DeshmukhFKS0J15} for discovering an execution that violates the specification. The falsification techniques may categorize counterexamples based on their robustness but they are geared towards generating their complete characterization.

In the domain of automated synthesis, the \emph{Counterexample Guided Inductive Synthesis} (CEGIS) framework~\cite{solar2008program} leverages counterexamples from verification for synthesis. The approach presented in this paper bears some resemblance to the CEGIS-based approach described in~\cite{raman2015reactive,RamanDMMSS14} where the verification condition that the system satisfies an STL specification is encoded as an MILP. If the specification is violated, one can investigate the results of the MILP to obtain counterexamples. In~\cite{ghosh2016diagnosis}, the authors extend the previous work and provide an intuition for the system failing to satisfy the specification. These work, in contrast to the presented work, do not compute reachable set for generating the counterexamples. A theoretical analysis of CEGIS based on different  counterexamples in the domain of program synthesis is attempted in~\cite{DBLP:journals/corr/JhaS14a}. It considers minimal and history bounded counterexamples which are aimed at localizing the error. The authors use these traces to investigate whether there are \emph{good mistakes} that could increase synthesis power and conclude that none of the two counterexamples are strictly good mistakes.

While model checking can find subtle errors, debugging in both hardware and program verification is shown to benefit from identifying different counterexamples or crucial failure sites. To obtain a minimal set of candidate error sites for all  counterexamples in an erroneous design, the work in ~\cite{1210088} introduces two greedy heuristics namely \emph{maximum pairwise distance} and \emph{efficient selection of error sites}. The  variations of a single counterexample are classified as positive and negative executions for analysis in~\cite{10.1007/3-540-44829-2_8}. It automates finding these executions and analyzing them for a  better summarized description of the erroneous elements.

The notion of causality is used to visually explain failure sites on the counterexample trace in~\cite{10.1007/978-3-642-02658-4_11}, whereas~\cite{10.1007/3-540-46002-0_31} focuses on inevitability towards the failure to capture more of the error in the error trace. The trace is presented as an alternation of fated and free segments: the fated segments show unavoidable progress towards the error while free segments show choices that, if avoided, may have prevented the error. Another technique \emph{delta debugging} ~\cite{10.5555/318773.318946} conducts binary search to discover and minimize difference between failing and succeeding runs of a program.

BDD's have been successfully used in computer aided design of logic circuits\cite{545618} and formal verification~\cite{1676819,Hu1995bddverify}. The application of BDDs to optimization is also growing in recent years~\cite{conf/cpaior/2011,Behle2007BinaryDD,doi:10.1137/1.9780898719789,dd-book-david-2016}. The total number of nodes in a BDD can grow exponentially large in the number of decision variables. Decision variables ordering plays a crucial role in determining the BDD size thus it is imperative to find an optimal ordering that minimizes the BDD size. However, the optimal ordering problem is shown to be computationally hard~\cite{10.1109/12.537122,10.1007/3-540-57568-5_270,10.1007/978-3-319-12691-3_33}. In practice, domain specific heuristics~\cite{10.1145/127601.127705,394067,10.5555/315149.315323,10.5555/1622420.1622423} are proposed for an efficient ordering. We apply a different heuristic which is based on our observation specifically about the overlap of the reachable set  with unsafe set.



\section{Preliminaries}
\label{sec:prelims}
We denote the set of Boolean values as $\mathbb{B}$.
States denoted as $\state{x}$ are the elements in $\reals^n$. The Inner product of two vectors $\state{v_1, v_2} \in \reals^n$ is denoted as $\state{v_1}^T \state{v_2}$. $\lvert\cdot\rvert$ operator denotes the cardinality of a given set. A linear constraint $\phi$ denotes a half-space in $\mathbb{R}^n$ is a mathematical expression $\state{a}^T \state{x} \leq \scalar{b}$, where $\state{a} \in \mathbb{R}^n$ is vector of coefficients and $\scalar{b} \in \mathbb{R} $ is a constant. Negation of a linear constraint $\phi \deq \state{a}^T \state{x} \leq \scalar{b}$ is another linear constraint $\neg \phi \deq  \state{(-a)}^T \state{x} \leq - (\scalar{b} + \epsilon)$ for $0 < \epsilon \ll 1$. In practice, whenever a solver gets $\phi \wedge \neg \phi$, the solution would be infeasible. Given a sequence $seq = s_1, s_2, \ldots$, the $i^{th}$
element in the sequence is denoted as $seq[i]$. Decision variables are elements in a set $\mathbb{B}$ of binary values 0 and 1.

\subsection{Discrete Time-invariant linear system}
\label{subsec:discrete-lti}

The behaviors of control systems with inputs are modeled with differential equations. In this work, we consider time-invariant affine systems with piece-wise bounded inputs.
\begin{definition}
\label{def:discreteLTIInputs}
An $n$-dimensional \emph{time-invariant affine system with bounded inputs} given as $\mathcal{H} \deq \tup{\mathcal{A}, \mathcal{B}}$ is denoted as $x_{t+1} = \mathcal{A} x_t + \mathcal{B}u_t$, $u_t \in \mathcal{U}$ where:
\begin{description}
\item[$\mathcal{A}$] $\in \reals^{n \times n}$ is a  time-invariant dynamics matrix,
\item[$\mathcal{B}$] $\in \reals^{n \times m}$ is a  time-invariant input  matrix, 
\item[$u_t$] $\reals_{\geq 0} \rightarrow \reals^m$ is the input function, assuming the system has $m$ inputs. $\mathcal{U} \subseteq \reals^{m \times 1}$ is the set of possible inputs.

\end{description}
\end{definition}

If $u_t$ is a constant function set to the value of $u$, then the system can be represented as $\tup{\mathcal{A}, \mathcal{B}u}$ which is a linear system with constant inputs.

Given an initial state $x_0$, a sequence of input vectors $\bar{u}$, the execution $\xi(x_0, \bar{u})$ of $\mathcal{H}$ is uniquely defined as the (possibly infinite) sequence $\xi(x_0, \bar{u}) \doteq \xi[0] \xrightarrow[]{\bar{u}[0]} \xi[1] \xrightarrow[]{\bar{u}[1]} \ldots$, where $\xi[t+1] = \mathcal{A} \xi[t] + \mathcal{B}\bar{u}[t], t \geq 0$ with $\xi[0] = \state{x_0}$ and $\xi[t] = \xi(\state{x_0}, \bar{u})[t]$ is the state at time $t$.

\begin{definition}
\label{def:reachability}
A state $\state{x}$ is \emph{reachable at $i^{th}$-step}, $0 \leq i$, if there exists an execution $\xi (\state{x_0}, \bar{u})$ of $\mathcal{H}$ such that $\xi(\state{x_0}, \bar{u})[i] = \state{x}$. Given an initial set $\Theta \subseteq \reals^n$, the set of possible inputs $\mathcal{U}$, the reachable set $Reach(\mathcal{H},\Theta,\mathcal{U})$ for such system is the set of all states that can be encountered by any $\xi(x_0, \bar{u})$ starting from any $x_0 \in \Theta$, for any valid sequence of input vectors $\bar{u}[t] \in \mathcal{U}, t \geq 0$.
The bounded time-variants of the execution and the reachable set, with time bound $\timebound \in \mathbb{N}$, are denoted by $\xi(\state{x_0}, \bar{u}, \timebound)$ and $\reachset$ respectively.  $Reach(\mathcal{H}, \Theta, \mathcal{U}, \timebound)[i]$ denotes the set of states reachable at a discrete time instance $i$.
\end{definition}

We now define the safety property for executions as well as for a set of possible inputs and a set of initial states.
\begin{definition}
\label{def:execSafe}
A given execution $\xi (\state{x_0},\bar{u})$ of system $\mathcal{H}$ is said to be safe with respect to a given unsafe set of states $\unsafe \subset \reals^n$ if and only 
if $\xi (\state{x_0},\bar{u})[t] \notin \unsafe, t \geq 0$. The state $\state{x_0}$ is called a \emph{counterexample} if $\exists t \geq 0$  and $\exists \bar{u}$ s.t. $\forall t' \leq t, \bar{u}[t'] \in \mathcal{U}$, we have $\xi (\state{x_0},\bar{u})[t] \in \unsafe$. Safety for bounded-time executions is defined similarly. 
\end{definition}

\begin{definition}
\label{def:reachSafe}
A discrete time-invariant linear system with bounded inputs $\mathcal{H}$ with initial set $\Theta$, set of possible inputs $\mathcal{U}$, time bound $\timebound$, and unsafe set $\unsafe$ is said to be safe with respect to its executions if $\reachset \cap \unsafe = \emptyset$. 
\end{definition}

%

%


\subsection{Reachable Set Computation}
\label{subsec:reach-set-compute}
We now present some building blocks in computation of the reachable set (from~\cite{bak2017rigorous,bak2017simulationinputs}). First is the generalized star representation used to symbolically represent the set of reachable states for potentially infinite state systems. Second is the algorithm for the reachable set computation.
%
%

\begin{definition}
\label{def:genStar}
A \emph{generalized star} (or simply star) $\Star$ is a tuple $\tup{\state{c},V,P}$ where $\state{c} \in \mathbb{R}^n$ is called the \emph{center}, $V = \{\state{v_1,v_2,\ldots, v_m}\}$ is a set 
of $m$ ($\leq n$) vectors in $\reals^n$ called the \emph{basis vectors}, and 
$P: \reals^n \rightarrow \{\top, \bot\}$ is a predicate.

A generalized star $\Star$ defines a subset of $\reals^n$ as follows.
\begin{eqnarray*}
\means{\Star} \deq \{ \state{x}\: |\: \exists {\bar{\alpha}} = [\bm{\alpha_1}, \ldots, \bm{\alpha_m}]^T \mbox{ such that } \\
\state{x} = \state{c} + \Sigma_{j=1}^m \bm{\alpha_j} \state{v_j} \mbox{ and } P(\bar{\alpha}) = \top \}
\end{eqnarray*}
Sometimes we will refer to both $\Star$ and $\means{\Star}$ as $\Star$. A star with  $P(\bm{\alpha}) = \bot$ is called as \emph{infeasible} which means the predicate $P$ does not have any satisfiable valuation of $\bar{\alpha}$. Additionally, we refer to the variables in $\bar{\alpha}$ as basis variables and the state variables $\state{x}$ as orthonormal variables. Given a valuation of the basis variables $\bar{\alpha}$, the corresponding orthonormal variables are denoted as $\state{x} = \state{c} + V \times \bar{\alpha}$. 
\end{definition}

\begin{remark}[Convex Predicates]
\label{rem:convex-stars}
We consider predicates $P$ which are conjunctions of linear constraints. That is, we consider stars which are convex sets. However, generalized star provides flexibility to compute the reachable set even when the initial set is non-convex~\cite{duggirala2016parsimonious}.
\end{remark}

\begin{definition}[Intersection of Stars]
Given two stars $\Star \doteq \tup{c, V, P}$ and $\Star' \doteq \tup{c, V, P'}$ with  center $c$ and the set of basis vectors $V$, the set intersection of these stars is a new star  $\bar{\Star} = S \cap S \doteq \tup{c, V, P \wedge P'}$. We abuse the set intersection operator to denote the intersection of 2 sets of predicates as well. For two sets $I \doteq \{P_1, P_2\}$ and $I' \doteq \{P_3\}$ where all predicates are represented in center $c$ and basis vectors $V$, we have $I \cap I' \doteq P_1 \wedge P_2 \wedge P_3$.
\end{definition}

\begin{definition}[Minkowski Sum with Stars]
\label{def:minkowski-sum-stars}
Given two stars $\Star = \tup{c, V, P}$ with $m$ basis vectors and $\Star' = \tup{c', V', P'}$ with $m'$ basis vectors, their Minkowski sum $\Star \oplus \Star'$ is a new star $\bar{\Star} = \tup{\bar{c}, \bar{V} , \bar{P}}$ with $m + m'$ basis vectors and (i) $\bar{c} = c + c'$, (ii) $\bar{V}$ is the list of
$m + m'$ vectors produced by joining the list of basis vectors of $\Star$ and $\Star'$, (iii)
$\bar{P}(\bar{\alpha}) = P (\bar{\alpha}_{m}) \wedge P'(\bar{\alpha}_{m'})$. Here $\bar{\alpha}_m \in \reals^m$ denotes the variables in $\Star$, $\bar{\alpha}_{m'} \in \reals^{m'}$
denotes the variables for $\Star'$, and $\bar{\alpha} \in \reals^{m+m'}$ denotes the variables for $\bar{\Star}$ (with
appropriate variable renaming).
\end{definition}

\noindent {\bf Reachable Set Computation Algorithm:} We briefly describe the technique for computing the reachable set. This is primarily done to present some crucial observations which will later be used in the algorithms for obtaining counterexamples. Longer explanation, proofs for  observations, and the algorithm to compute simulation-equivalent reachable set for linear continuous dynamical affine systems with bounded inputs are available in prior works~\cite{duggirala2016parsimonious,bak2017rigorous,bak2017simulationinputs}.

At its crux, the algorithm exploits the superposition principle of linear systems and computes the set of reachable states using the  generalized star representation. 
%
%
The algorithm yields the reachable set $Reach(\mathcal{H},\Theta,\mathcal{U},\timebound)$ as a sequence of stars $\Star_0, \Star_1, \Star_2, \ldots, \Star_T$ such that $\Star_0 = \Theta$ and for $i \in[0,T-1]$, 
\begin{align}
\label{eqn:starwinput}
\Star_{i+1} &\doteq \mathcal{A}\Star_{i} \oplus  \mathcal{BU} \nonumber \\
&\doteq \mathcal{A}^{i+1}\Star_0 \oplus \mathcal{A}^{i}\mathcal{BU}\oplus \mathcal{A}^{i-1}\mathcal{BU}\oplus \ldots \oplus \mathcal{BU}.
\end{align}

The first term $A^{i+1}S_0$ is the reachable set of the system without inputs and the remainder of the terms characterize the effect of inputs. Consider the $j^{th}$ term in the remainder, namely, $\mathcal{A}^j\mathcal{BU}$ in Equation~\ref{eqn:starwinput}. This term
is exactly same as the reachable set of states starting from an initial set $\mathcal{BU}$ after $j$ time units evolving according to the dynamics $x_{t+1} = \mathcal{A}x_t$.

Furthermore, the set $\mathcal{BU}$ can be represented as a star $\tup{c, V, P}$ with $m$ basis vectors, for an $n$-dimensional system with $m$ inputs. This is done by taking the origin as the center $c$, the set $\mathcal{B}$ as the star’s $n \times m$ basis
matrix $V$ , and using the linear constraints $\mathcal{U}$ as the predicate $P$ , replacing each input $u_i$ with $\alpha_i$. Now, the effect of the inputs after $j$ time units is computed as the Minkowski sum of stars denoted as $\mathcal{U}_j \oplus \mathcal{U}_{j-1} \oplus \mathcal{U}_{j-2} \oplus \ldots \oplus \mathcal{U}_0$ where $\mathcal{U}_j$ is the effect of inputs at $j^{th}$ time step and $\mathcal{U}_0 \doteq \mathcal{U}$.

Notice that both the number of variables in the star and the number of constraints grow with each Minkowski sum operation. Since we focus on bounded piece-wise constant inputs, for $m$ inputs, the number of variables (or basis vectors) grow by $m$ and the number of constraints grow by $2m$ after every unit time step to incorporate the effect of inputs. For instance, the set of states reachable exactly after $j$ time steps is denoted as $S_j$ which would have $n + (j \times m)$ basis vectors and $|P_0| + (j \times 2 m)$ constraints where $|P_0|$ is the number of constraints in the initial star $S_0$. If the initial set is given as an $n$-dimensional hyper-rectangle then $|P_0| = 2n$.

For a time-invariant affine system with constant (including 0) input $u$, the reachable set computation algorithm would yield the reachable set as a sequence of stars $\Star_0, \Star_1, \Star_2, \ldots, \Star_T$ such that $\Star_0 = \Theta$ and for $i \in[0,T-1]$, 
\begin{align}
\Star_{i+1} &\doteq \mathcal{A}\Star_{i} + \mathcal{B}u \nonumber \\
&\doteq \mathcal{A}^{i+1}\Star_0 + \mathcal{A}^{i}\mathcal{B}u+\mathcal{A}^{i-1}\mathcal{B}u+\ldots+\mathcal{B}u.
\end{align}

\textbf{Observation:} Given a star $\Star_i \doteq \tup{\state{c_i}, V_i, P} \in \reachset$ and a valuation $\bar{\alpha}$ of the basis variables such that  $P(\bar{\alpha}) = \top$, one can use this valuation of basis variables to generate the trace starting from the initial set $\Theta$ to $\Star_{i}$. For a valuation $\bar{\alpha}$ such that $P(\bar{\alpha}) = \top$, the corresponding bounded time execution (for time bound $\timebound$) would be $\state{c_0} + V_0 \times \bar{\alpha}, \state{c_1} + V_1 \times \bar{\alpha}, \ldots, \state{c}_{\timebound} + V_{\timebound} \times \bar{\alpha}, \ldots$. This observation is crucial for our algorithm as it reduces the problem of finding the counterexample from the functional space of executions $\xi$ to the valuation of the basis variables $\bar{\alpha}$; each valuation $\bar{\alpha}$ designates a unique execution.


\subsection{Counterexamples and Constraint Propagation}
\label{subsec:constr-prop}
In order to obtain the set of executions that reach the unsafe set $\unsafe$ at time step $i$, we need to find the valuations of basis variables $\bar{\alpha}$ such that $P(\bar{\alpha}) = \top$ and $\left(\state{c_i} + V_i \times \bar{\alpha}\right) \in \unsafe$. We begin with computing the reachable set as a sequence of stars $\Star_0, \Star_1, \ldots, \Star_\mathrm{T}$ where $\Star_0 \doteq \Theta$, $\Star_i \deq \tup{\state{c}_i, V_i, P} \in \reachset$. Now, given star $\Star_i$, we can represent $\unsafe$ as another star $\tup{\state{c_i}, V_i, P_i^{\unsafe}}$ by converting each constraint $(\state{a^T x} \leq \scalar{b}) \in \unsafe$ as $\state{a^T} (\state{c_i} + V_i \bar{\alpha}) \leq \scalar{b}$. 
This gives us $P_i^{\unsafe} \doteq A^T (\state{c_i} + V_i \bar{\alpha}) \leq \state{b}$ where $A \in \reals^{(n + (i \times m)) \times j}, \state{b} \in \reals^j$, and $j$ is the number of constraints in $P_i^{\unsafe}$. 
We next check feasibility of the predicate $P \wedge P_i^{\unsafe}$. There is no counterexample reaching $\unsafe$ at time $i$ if $\left(P \wedge P_i^{\unsafe}\right) (\bar{\alpha}) = \bot$.
Otherwise, we make use of the previous observation to propagate constraints so that all the executions that reach the unsafe region $\Star_i \cap \unsafe \doteq \langle \state{c_i}, V_{i}, P \wedge P_i^{\unsafe} \rangle$ at time step $i$ would originate from the set $\Theta_i \doteq \langle \state{c_0}, V_0, P \wedge  P_i^{\unsafe} \rangle \subseteq \Theta$. In other words, the predicate $P \wedge P_i^{\unsafe}$ denotes the set of $\bar{\alpha}$ characterizations for the counterexamples reaching $\unsafe$ at time step $i$. 
Considering $P \doteq (C^T \state{x} \leq \state{d})$, we solve a system of constraints $\left(A^T (\state{c_0} + V_0 \bar{\alpha}) \leq \state{b} ~\wedge~ C^T (\state{c_0} + V_0 \bar{\alpha}) \leq \state{d}\right)$ to find a satisfiable valuation $\bar{\alpha}$. Then we generate a  desired counterexample as an execution $\state{c_0} + V_0 \times \bar{\alpha}, \state{c_1} + V_1 \times \bar{\alpha}, \ldots$ where $\left(\state{c_i} + V_i \times \bar{\alpha}\right) \in \unsafe$.

The extension of this approach to multiple time steps is straightforward. The executions that reach $\unsafe$ at two different time steps $i$ and $i'$ would originate from $(\Theta_i \cap \Theta_{i'}) \doteq \langle \state{c}, V, P \wedge  P_i^{\unsafe} \wedge P_{i'}^{\unsafe} \rangle \subseteq \Theta$. We can solve the corresponding system of constraints to obtain a valuation $\bar{\alpha}$ and generate a desired counterexample. 

\begin{remark}
\label{rem:general-u}
For computing the characterization of counterexamples, we only consider time steps (or star indices) at which the reachable set overlaps with the unsafe set $\unsafe$. Further, although we focus on safety specification in this paper as we are dealing with safety critical systems, $\unsafe$ symbolizes the violation of a general performance specification.
\end{remark}

\subsection{Metric Temporal Logic}

Metric Temporal Logic (MTL) is defined over a finite set  $\mathcal{P}$ of atomic propositions. Each proposition $p \in \mathcal{P}$ at discrete time $t \in \mathbb{N}$ takes a value from the boolean set $\{\top,\bot\}$. A timed word is defined as $\sigma : \mathbb{N} \to 2^{\mathcal{P}}$, where $\sigma[t] \in 2^{\mathcal{P}}$ is
the set of propositions that are true at time t~\cite{7799150}. The syntax of
MTL formulas is recursively defined as:

\begin{equation}
\label{eqn:mtlformula}
\varphi ::= \top ~|~ p ~|~ \neg \varphi ~| ~\varphi_1 \wedge \varphi_2~|~ \varphi_1 \mathrm{U}_I \varphi_2
\end{equation}

where $\neg$ and $\wedge$ are boolean
negation and conjunction operators, respectively, and $\mathrm{U}$ is the timed \emph{until} operator with a time interval $I \subseteq [i_1, i_2]$ where $0 \leq i_1 \leq i_2 \leq \timebound$ and $\timebound \in \mathbb{N}^+$ is the time bound. Other temporal operators are
constructed using the syntax above. The temporal
finally (\emph{eventually}) is defined as $\mathrm{F}_{I} \varphi := \top \mathrm{U}_I \varphi$ and temporal globally
(\emph{always}): $\mathrm{G}_I \varphi := \neg \mathrm{F}_I \neg \varphi$.
Word $\sigma$ satisfies MTL formula
$\varphi$, denoted by $\sigma \models \varphi$, if $\sigma_0 \models \varphi$, where $\sigma_0$ is a timed sequence $``\sigma[0], \sigma[1], \ldots"$ starting at time $0$. The language of $\varphi$ is
the set of all words satisfying $\varphi$. The semantics for MTL are inductively defined and can be referred in~\cite{THATI2005145}.

We consider specifications described using MTL where
each of its atomic propositions is over a set of linear
constraints:
$$
p := A_p \state{x} \leq \state{b}_p
$$

where $A_p \in \reals^{|p| \times n}$, $\state{b}_p \in \reals^{|p|}$, and $|p|$ is the number of constraints in proposition $p$. We restrict our attention to the unsafe set $\Psi$ specified as a linear constraint $\state{a^T x} \leq \scalar{b}$ such that it divides the state space in two half spaces.

\section{Complete Characterization of Counterexamples}
\label{sec:characterize-problem}
We begin with a simple example that helps us in building the set up towards problem definition, proposed solution, BDD  reduction and related discussion.

\begin{example}
\label{example:osc-particle} \emph{Oscillating Particle}~\cite{10.1007/s10703-011-0118-0} is a 3-dimensional discrete time-invariant linear system,  $\mathcal{H} \deq \tup{\mathcal{A}, \mathcal{B}}$, where:
$$
\mathcal{A} = \begin{pmatrix}
0.722 & -0.523 & 0 \\
0.785 & 0.696 & 0\\
0 & 0 & 0.931 \\
\end{pmatrix}, 
\mathcal{B} = \begin{pmatrix}
0 \\
0.1\\
0.1\\
\end{pmatrix}. \\
$$
\end{example}

Given $\Theta \doteq \left([-0.1, 0.1], [-0.8, -0.4], [-1.07, -1]\right)$, $\mathcal{U} \doteq [-0.04, 0.04]$ and $\timebound=15$, the reachable set $\reachset$ computed in the form of generalized stars by a linear hybrid system reachability analysis tool, HyLAA~\cite{duggirala2016parsimonious} is shown in Fig.~\ref{fig:osc-reach-set}. 

The safety specification is $\phi \doteq \mathrm{G}_{[0, 15]} \neg p$, where $p = y \geq 0.4$ is an atomic proposition in $\mathcal{P}$. The system violates $\phi$ because $\reachset$ has a  non-empty intersection with the unsafe set $\unsafe \doteq p$. Fig.~\ref{fig:intro-reg-control} emphasizes that different system executions can overshoot the threshold at multiple (possibly different) time instances. The reachable set illustration would give a much better idea of the modalities of these executions instead of explicitly identifying individual overshoot profiles in a likely infinite state system.

\subsection{Introducing Characterization}
\label{subsec:complete-characterize}
Suppose the reachable set for a given linear dynamical system and safety specification violates the specification (i.e., overlaps with the unsafe set $\unsafe$) at $k$ (not necessarily contiguous) time steps. The \emph{characterization} of an execution $\xi (\state{x_0},\bar{u})$ is defined as a function $\mathcal{C}_{\unsafe}: \left(\reals^n\right)^T \rightarrow \left(2^{\mathcal{P}}\right)^k$, which is a projection of $\xi (\state{x_0}, \bar{u})$ into the space of propositions defining the unsafe set. Thus the \emph{complete characterization} is the set of all unique characterizations of a safety violation.
$$
\means{\mathcal{C}_{\unsafe}} = \{ \mathcal{C}_{\unsafe} (\xi (\state{x_0}, \bar{u})) \: | \: \xi (\state{x_0},\bar{u}) \in \reachset \}.
$$

That is, $\means{\mathcal{C}_{\unsafe}}$ is set of all system  executions mapped into the space $\left(2^{\mathcal{P}}\right)^k$ for the unsafe set. 
%
Sometimes we refer to both $\mathcal{C}_{\unsafe}$ and $\means{\mathcal{C}_{\unsafe}}$ as $\mathcal{C}_{\unsafe}$.\\

\textbf{Problem Statement I:} Given a linear system $\mathcal{H}$, initial set $\Theta$, a set of possible inputs $\mathcal{U}$, a safety specification $\phi_I^{\unsafe}$ denoted as an MTL formula defined over a set $\unsafe$ and time interval $I \doteq [i_1,i_2]$, if $\exists i' \in I$ such that $\reachset[i'] \cap \unsafe \neq \emptyset$, compute $\means{\mathcal{C}_{\unsafe}}$.


\begin{figure}[ht]
\centering
\includegraphics[width=88mm]{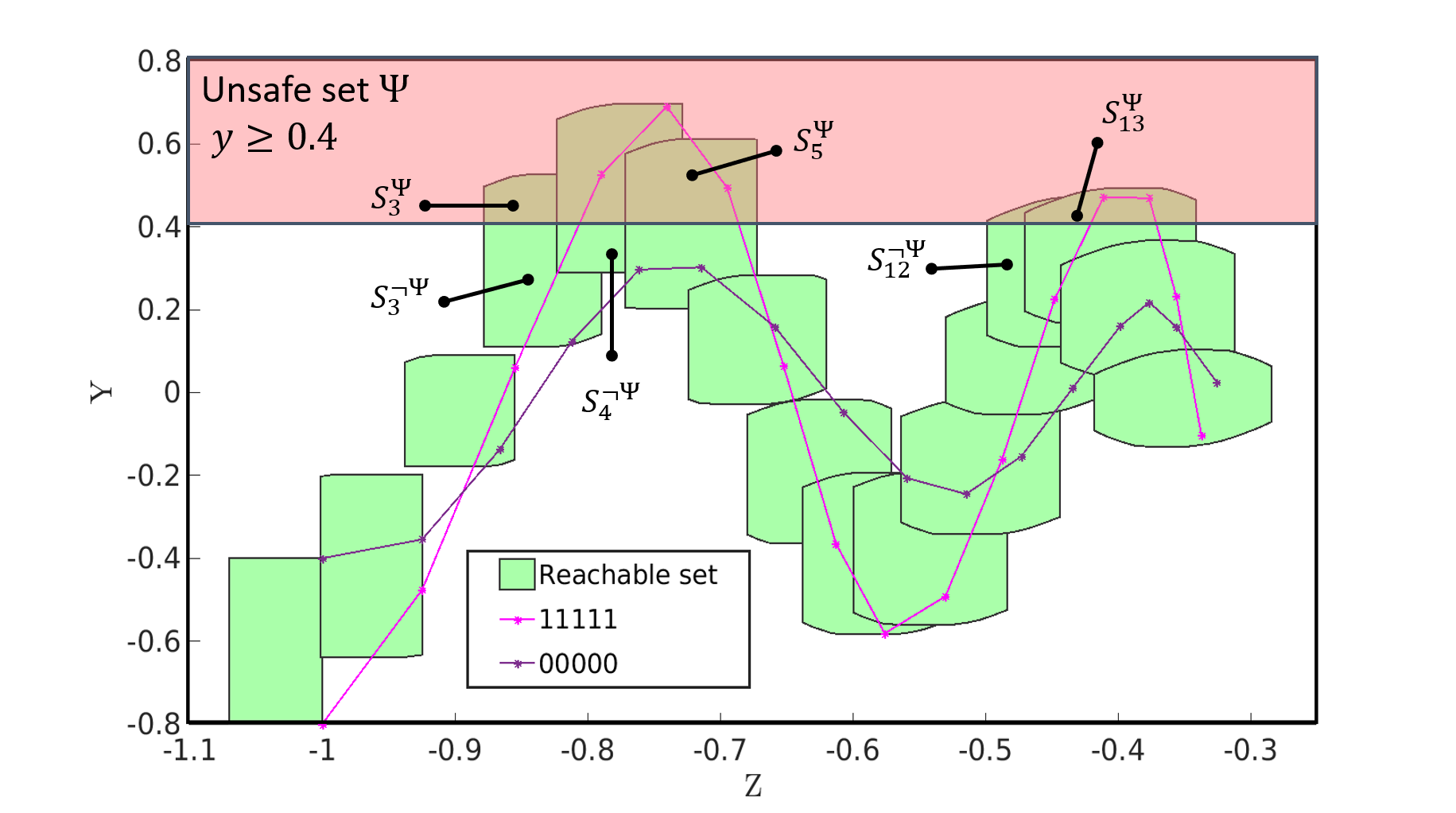} 
\caption{\small{Reachable set computed in HyLAA for Example~\ref{example:osc-particle}. The system violates the safety specification at $5$ time steps. The execution labeled as $11111$ denotes a characterization for which the reachable set has a non-empty overlap with $\unsafe$  at all $5$ time steps. Whereas, the execution labeled as $00000$ is a characterization for which the reachable set does not overlap with $\unsafe$ at all.}}
\label{fig:osc-reach-set}
\end{figure}

Fig.~\ref{fig:osc-reach-set} exhibits that the reachable set overlaps with $\unsafe$ (i.e., it satisfies proposition $p$) at $5$ steps during time interval $[0, 15]$. The overlap specifically occurs at the following time steps: $3^{rd}, 4^{th}, 5^{th}, 12^{th}$ and $13^{th}$. In the figure, one characterization of this safety violation denotes system executions that reach unsafe set $\unsafe$ at all these $5$ time steps. Another characterization represents executions that do not overlap with the unsafe set at all. After having obtained these characterization, a question may arise whether there are executions that violate safety at only $3^{rd}$ time step, at only $12^{th}$ time step or both. The brute force way of obtaining such characterizations is to compute all binary strings of length $5$, and check for the existence of a  counterexample corresponding to each string. However, computing these exponential number of strings is not desirable by any practical means. Thus we would present a computationally efficient approach to represent complete characterization of these counterexamples.

\subsection{Constraint propagation demonstration}
Before we introduce the approach to obtain complete characterization of a violation of a safety specification $\phi_I^{\unsafe}$, we demonstrate constraint propagation step in our running example.
  We define an \emph{ordered} set $\Pi^{\unsafe} \deq \{S_i \cap \unsafe \: |\: S_i \in \reachset, i \in I, S_i \cap \unsafe \neq \emptyset\}$. So, we have $\Pi^{\unsafe} = \{S_3\cap \unsafe, S_4\cap \unsafe, S_5\cap \unsafe, S_{12}\cap \unsafe, S_{13}\cap \unsafe\}$. Or, we simply write $\Pi = \{S_3^{\unsafe}, S_4^{\unsafe}, S_5^{\unsafe}, S_{12}^{\unsafe}, S_{13}^{\unsafe}\}$ where $S_i^{\unsafe} = S_i \cap \unsafe$. We define $\Pi^{\neg \unsafe}$ in a similar manner to obtain another ordered set $\Pi^{\neg \unsafe} = \{S_3^{\neg \unsafe}, S_4^{\neg \unsafe}, S_5^{\neg \unsafe},  S_{12}^{\neg \unsafe}, S_{13}^{\neg \unsafe}\}$ where $S_i^{\neg \unsafe} = \ S_i \cap \neg \unsafe$. These stars and their predicates denote valuations of the proposition in given temporal logic formula $\phi_I^{\unsafe}$. Further, we only need to propagate constraints of the stars that are unsafe as explained in Remark~\ref{rem:general-u} in  Section~\ref{subsec:constr-prop}.

By using the technique described in Section~\ref{subsec:constr-prop}, we propagate the constraints for elements in $\Pi^{\unsafe}$ and $\Pi^{\neg \unsafe}$ to the initial set $\Theta$; this results into sets $\Theta_{\Pi}^{\unsafe}$ and $\Theta_{\Pi}^{\unsafe}$ respectively. Any execution that originates from $\Theta_{\Pi}^{\unsafe}[j] \doteq \langle \state{c_0}, V_0, P^j \rangle \subseteq \Theta$ would reach $\Pi^{\unsafe}[j]$ (i.e., $S_i^{\unsafe}$) at $i^{th}$ time step, where $P^j \doteq P \wedge  P_i^{\unsafe}$. \emph{Note that superscript $j$ indexes an element in the ordered set $\Theta_{\Pi}^{\unsafe}$ while subscript $i$ is the time step at which any execution starting from this $j^{th}$ element reaches the unsafe set.} 
Similarly, an execution starting from $\Theta_{\Pi}^{\neg \unsafe}[j] \doteq \langle \state{c_0}, V_0, \bar{P}^j \rangle \subseteq \Theta$ would reach $\Pi^{\neg \unsafe}[j]$ (i.e., $ S_i^{\neg \unsafe}$) at $i^{th}$ time step, where $\bar{P}^j \doteq P \wedge  P_i^{\neg \unsafe}$. 

\begin{remark}
\label{rem:covex-predicate}
It is important to note here that we only focus on the unsafe set $\unsafe$ given as a linear constraint $a^T x \leq \scalar{b}$, thus the predicates $P^j$ and $\bar{P}^j$ have the same set of constraints except the one corresponding to $\Psi$ and $\neg \Psi$ respectively. In other words, $P^j$ and $\bar{P}^j$ denote two half spaces separated by the hyperplane $a^Tx = \scalar{b}$. Consequently, the predicate $P^{\neg \unsafe}$ or $\bar{P}^j$ is a conjunction of linear constraints.
\end{remark}

\begin{figure}[ht]
\centering
\includegraphics[width=60mm,height=40mm]{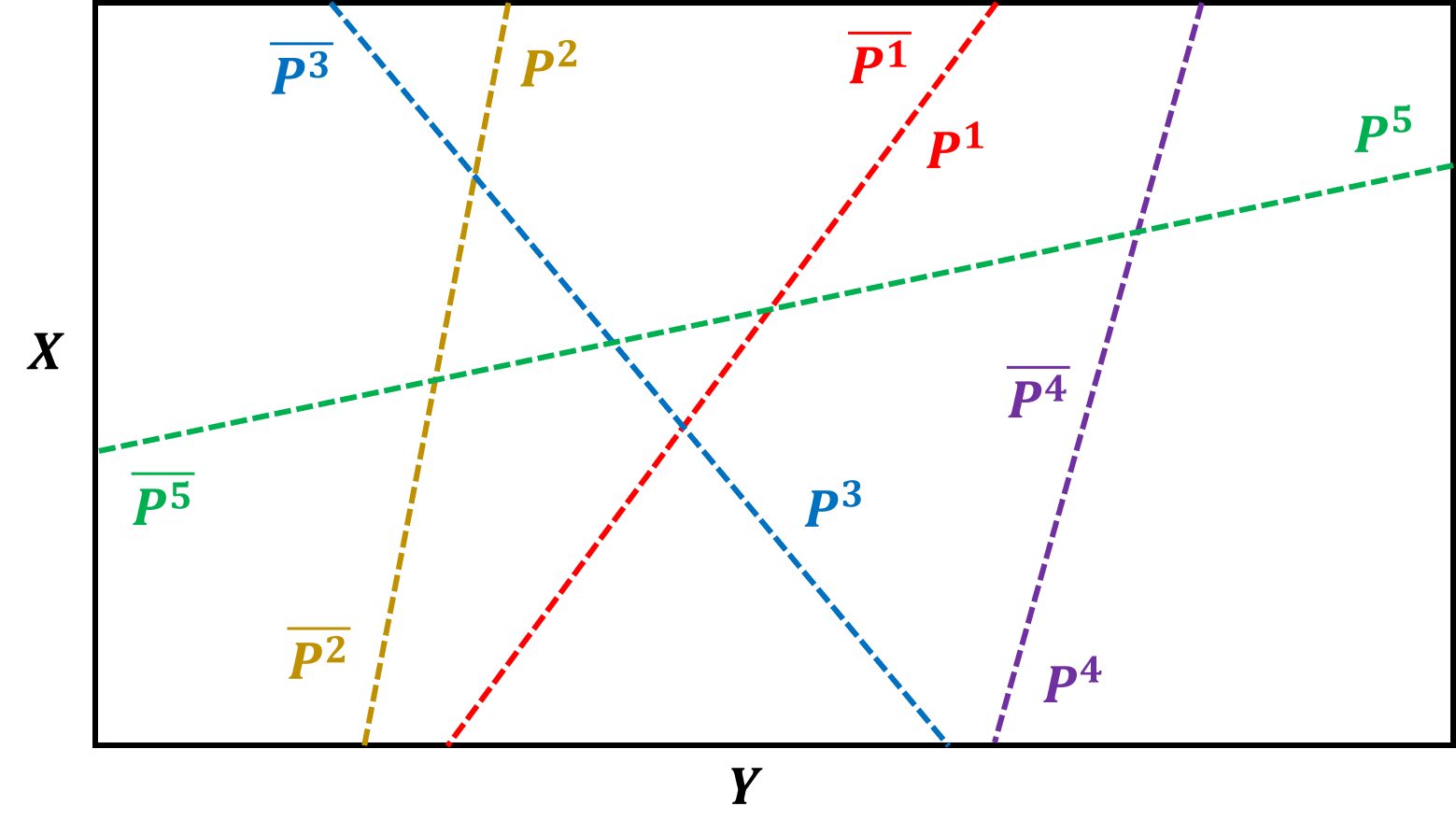}
\caption{\small{Initial set $\Theta$ after constraint propagation. $P^j \doteq P \wedge P^{\unsafe}_i$, where $j$ indexes an element in the ordered set $\Theta_\Pi^{\unsafe}$ while $i$ is the time step at which any execution starting from this $j^{th}$ element reaches the unsafe set $\unsafe$. Similarly, $\bar{P}^j \doteq P \wedge P^{\neg \unsafe}_i$, where $j$ indexes an element in $\Theta_{\Pi}^{\unsafe}$ such that any execution originating from this element does not enter  $\unsafe$ at time step $i$.}}
\label{fig:theta-constr-prop}
\end{figure}

%
It easily follows that $\forall j, \Theta_{\Pi}^{\unsafe}[j] \cap \Theta_{\Pi}^{\neg \unsafe}[j] = \emptyset$ and $\Theta_{\Pi}^{\unsafe}[j] \cup \Theta_{\Pi}^{\neg \unsafe}[j] = \Theta$. The elements in $\Theta_{\Pi}^{\unsafe}$ and $\Theta_{\Pi}^{\neg \unsafe}$  are specified in the space of same center $\state{c_0}$ and basis vectors $V_0$ of  $\Theta$. Therefore, for the ease of exposition, we can refer to the star $\Theta_{\Pi}^{\unsafe}[j]$ by its predicate $P^j$ and star $\Theta_{\Pi}^{\neg \unsafe}[j]$ by $\bar{P}^j$ (which is a conjunction of linear constraints). 
%
The resultant of constraint propagation step is shown in Fig.~\ref{fig:theta-constr-prop}. 

Now a system execution that visits the unsafe set $\unsafe$ at just  $3^{rd}, 4^{th}, 5^{th}$ time steps can be obtained by finding a satisfiable valuation  of $\bar{\alpha}$ for the predicate $\left(P^1 \wedge P^2 \wedge P^3 \wedge \bar{P}^4 \wedge \bar{P}^5\right) (\bar{\alpha})$. The characterization of such counterexample is $\{\{p\}, \{p\}, \{p\}, \{\neg p\}, \{\neg p\}\}$ or $\{p, p, p, \neg p, \neg p\}$ or $\{\top, \top, \top, \bot, \bot\}$ or $\{1, 1, 1, 0, 0\}$. Similarly, the characterization of a counterexample that reaches $U$ at only $4^{th}$ time instance is $\{0, 1, 0, 0, 0\}$, and the characterization of a longest counterexample is $\{1, 1, 1, 1, 1\}$.
%
A few other characterizations of this safety violation are $\{1, 1, 0, 0, 0\}$, $\{0, 1, 1, 0, 1\}$, and $\{0, 0, 1, 0, 0\}$ etc. A characterization is \emph{valid} if associated system of predicates is \emph{feasible}. For example, the characterization $\{1, 1, 1, 0, 1\}$ is \emph{valid} because $\left(P^1 \wedge P^2 \wedge P^3 \wedge \bar{P}^4 \wedge P^5\right) (\bar{\alpha}) = \top$. But 
a characterization $\{1, 0, ?, ?, ?\}$ is not valid because $\left(P^1 \wedge \bar{P}^2\right) (\bar{\alpha}) = \bot$. 

These multiple characterizations can provide some insights behind a safety violation thus assisting a control designer in state space exploration. Nonetheless, the main challenge remains is the combinatorial nature of the problem. The strings can be exponential in the number of times the reachable set enters the unsafe set. In a general case, there are $\left(2^{m}\right)^k$ binary strings for a violation of the specification with $m$ propositions. But computing and checking validity of exponential number of strings is both computationally inefficient and practically undesirable. 
A binary decision diagram (BDD) yields a tractable and efficient representation of binary functions and strings. Therefore we use a decision diagram for generating complete characterization of counterexamples.



\subsection{Binary Decision Diagram}
\label{subsec:bdd}

A \emph{Binary Decision Diagram} (BDD)~\cite{bdd_78} is a graphical data structure for representing a boolean function. It compactly represents a set of satisfiable assignments (solutions) to decision variables $z_i$ for a given boolean function $f : \mathbb{B}^n \rightarrow \mathbb{B}$ i.e.,
$$
f (z_1, z_2, \ldots, z_n) \rightarrow \{0, 1\}, z_i \in \{0, 1\}.
$$
It is typically represented as a \emph{rooted}, \emph{directed} and \emph{acyclic} graph which consists of several (variable) nodes and two terminal (valued) nodes namely \texttt{False(0)} and \texttt{True(1)}. Each variable node is labeled by a decision  variable $z_i$ and has two children - \emph{low} and \emph{high}. The directed edge from  variable node $z_i$ to $low$ represents an assignment $0$ to $z_i$ and the edge to $high$ denotes the assignment $1$.

\begin{definition}
\label{def:obdd}
A BDD is said to be \emph{ordered} (OBDD) if the variables have fixed order along all paths from  root to a terminal node in the graph.
\end{definition}
\begin{definition}
\label{def:rbdd}
A BDD is called \emph{reduced} (RBDD) if following two rules have been applied to the graph.
\begin{itemize}
\item \textbf{Isomorphism:} If two variable nodes are labeled by the same decision variable and have the same set of paths starting from them (to the terminal nodes), then these nodes are called \emph{isomorphic}. One of them is removed and the incoming edges to the removed node are redirected to its isomorphic node.
\item \textbf{Elimination:} If a variable node $\eta$ has isomorphic children, then $\eta$ is removed and incoming edges to $\eta$ are redirected to either of its children.
\end{itemize}
\end{definition}

A BDD which is both \emph{ordered} and \emph{reduced} is called Reduced Ordered Binary Decision Diagram (ROBDD). We only apply \emph{isomorphism} rule to our OBDD. The term BDD refers to OBDD (not necessarily reduced) in this paper. In an ordered BDD, each non-terminal level is associated with only one decision variable. The size of a diagram is measured in its total number of nodes ($\mathcal{N}$) and its width ($\mathcal{W}$) which is defined as the maximum number of nodes at any level.

\begin{definition}[Ordering]
For a set $\mathbb{N}^k \doteq \{1,2,\ldots,k\}$, an ordering $O: \mathbb{N}^k \rightarrow \mathbb{N}^k$ is an injective mapping such that for any $j \in \mathbb{N}^k$, there is $j' \in \mathbb{N}^k$ such that $O(j') = j$.
The operation of applying an ordering $O$ to a set $\Delta$ of cardinality $k$ is called  \algnm{\algfnt{enumerate}} which is defined as a function $\Delta \times O \rightarrow \Delta$ such that for any $\delta_j \in \Delta$, there is $\delta_{j'} \in \Delta$, $j, j' \in \{1, 2, \ldots, k\}$, such that $O(j') = j$. 
\end{definition}

Decision variable ordering plays a key role in determining the size of a decision diagram, which is observed in our evaluation results as well.


\section {Computing complete characterization}

\begin{algorithm}
\SetKwInOut{Input}{input}\SetKwInOut{Output}{output}\SetKw{Return}{return}
\Input{Initial set: $\Theta$, reachable set: $\reachset$, safety specification: $\phi_I^{\Psi}$, ordering: $O$}
\Output{BDD: $\mathcal{G}$, its node count $\mathcal{N}$ and width $\mathcal{W}$, set of valid paths: $\mathcal{C}_{\unsafe}$ }
$\Pi^{\unsafe} \deq \{S_i \cap \unsafe 
\:|\: S_i \in \reachset, i \in I, S_i \cap \unsafe \neq \emptyset\}$\;
$\Pi^{\neg \unsafe} \deq \{S_i \cap \neg \unsafe \: |\: S_i \in \reachset, i\in I\}$\;
$\Pi^{\unsafe} \gets \algnm{\algfnt{enumerate}} (\Pi^{\unsafe}, O)$\label{alg1:ln-enumerate1}\; 
$\Pi^{\neg \unsafe} \gets \algnm{\algfnt{enumerate}} (\Pi^{\neg \unsafe}, O)$\label{alg1:ln-enumerate2}\;
$\Theta_{\Pi}^{\unsafe}, \Theta_{\Pi}^{\unsafe} \gets \algnm{\algfnt{propagate\_constraints}}(\Pi^{\unsafe}, \Pi^{\neg \unsafe}, \Theta)$\label{alg1:ln-const-prop}\;
$\mathcal{G} \gets \algnm{\algfnt{init\_bdd}}(\Theta)$\label{alg1:init-bdd}\tcp*{creates root, terminals}
$\mathcal{N} \gets 1$\tcp*{for $root$ node}
$\mathcal{V} \gets [\mathcal{G}.root]$\tcp*{$root$ is at level 0}
$\mathcal{N} \gets \mathcal{N} + 2, \mathcal{W} \gets 2$\tcp*{for terminal nodes $t_0,  t_1$}
$k \gets |\Theta_{\Pi}|$\tcp*{\# of decision variables}
\For{$1 \leq j \leq k$}{\label{ln:begouterloop}
    $\mathcal{V}' \gets \emptyset$\tcp*{set of nodes at level $j$}
    \For{$ \eta \in \mathcal{V}$}{\label{ln:beginnerloop}
        $\mathcal{V}' \gets \mathcal{G}.\algnm{\algfnt{process\_node}}(\eta, j, \mathcal{V}', \Theta_{\Pi}^{\unsafe}, \Theta_{\Pi}^{\neg \unsafe})$\;
    }\label{ln:endinnerloop}
    $\mathcal{N} \gets \mathcal{N} + |\mathcal{V}'|$\label{alg1:udpate-node-count}\tcp*{update nodes count}
    \If{$|\mathcal{V}'| > \mathcal{W}$}{ 
    $\mathcal{W} \gets |\mathcal{V}'|$\label{alg1:udpate-width} \tcp*{update width}
    }
    $\mathcal{V} \gets \mathcal{V}'$\label{alg1:update-v}\tcp*{progress to next level}
}\label{ln:endouterloop}
$\mathcal{C}_{\unsafe} \gets \algnm{\algfnt{traverse}}(\mathcal{G}.root, \lambda, $`'$, \emptyset)$\label{alg1:traverse-bdd}\tcp*{$\lambda$ is  empty string}

{\bf return} $(\mathcal{G}, \mathcal{N}, \mathcal{W}, \mathcal{C}_{\unsafe})$\;
\caption{\small \algnm{\algfnt{construct\_bdd}} algorithm. The algorithm - enumerates the unsafe elements in the reachable set as per the given ordering $O$, propagates constraints, initializes BDD $\mathcal{G}$ and populates its nodes in \algnm{\algfnt{process\_node}}, and finally traverses the BDD to compute the set $C_{\unsafe}$ of valid characterizations.}
\label{alg:construct-bdd}
\end{algorithm}

\subsection{BDD construction}
We first describe the basic idea. At each level (from root to terminal), a predicate (say $P'$) is selected based on the the order that is pre-determined by $O$. It is then assessed whether $P'$  renders the predicate (say $P_{\eta}$) of every individual node $\eta$  at that level (in-)feasible. If the predicate ($P_{\eta} \wedge P'$) is evaluated to be feasible, it is added as the $high$ child of $\eta$; otherwise terminal $0$ is assigned as its $high$ child. The same process is repeated with predicate $\bar{P'}$ for determining $low$ child of $\eta$.
Selection of predicate $P'$ is analogous to assigning $1$ to corresponding decision variable at that level while $\bar{P'}$ reflects the valuation $0$ of the decision variable. We maintain two sets $\Pi^{\unsafe}$ and $\Pi^{\neg \unsafe}$ for predicates and their negative counterparts. The union of all these (non-terminal) children computed at a particular level constitute the set of nodes for  next iteration. The procedure terminates after $k$ iterations where $k$ is the number of decision variables. Once BDD construction is completed, it enumerates the paths (from the beginning) in the diagram for generating all characterizations of the safety violation.

The algorithm for computing characterization of a safety violation can be divided into 3 main routines. 

\algnm{\algfnt{1. construct\_bdd}}: Algorithm~\ref{alg:construct-bdd} computes the sets $\Pi^{\unsafe}$ and $\Pi^{\neg \unsafe}$, enumerates their elements as per the given ordering $O$, and propagates constraints (lines~2-5). As a consequence of constraint propagation, the predicates of the elements in $\Theta_{\Pi}^{\unsafe}$ and $\Theta_{\Pi}^{\unsafe}$ are now specified in center $\state{c_0}$ and basis vectors $V_0$. So a reference to a star or its predicate in this section implicitly considers $\state{c_0}$ as the center and $V_0$ as basis vectors. Further, we use the terms node and star interchangeably because each BDD node denotes a star. 

%

The initialization step of BDD $\mathcal{G}$ (line~\ref{alg1:init-bdd}) creates its \emph{root} node (i.e., star $\langle \state{c_0}, V_0, P \rangle$) in addition to the terminal nodes $t_0$ (i.e., star $\langle \state{c_0}, V_0, \bot \rangle$) and $t_1$ (i.e., star $\langle \state{c_0}, V_0, \top \rangle$). The diagram is constructed in the breadth-first manner where \emph{root} is the only node at level $0$.
 $\mathcal{V}$ denotes the set of nodes at the current level (starting from $0$), and $\mathcal{V'}$ maintains nodes at the next level. In each iteration of the outer loop (lines~\ref{ln:begouterloop} -~\ref{ln:endouterloop}), the children of every node $u \in \mathcal{V}$ are computed via \algnm{\algfnt{process\_node}}, and added to the set $\mathcal{V'}$. After processing all of the nodes in $\mathcal{V}$, nodes count $\mathcal{N}$ and width $\mathcal{W}$ are updated (lines~\ref{alg1:udpate-node-count} and~\ref{alg1:udpate-width}). Additionally, $\mathcal{V}$ is  replaced with $\mathcal{V'}$ (line~\ref{alg1:update-v}) for the next iteration.
%
 


\algnm{\algfnt{2. process\_node}}: Algorithm~\ref{alg:process-node} computes $high$ and $low$ children of a node $\eta$ at level $(j-1)$ where $j \in [1, k]$. We now discuss the technique for computing the $high$ child. Initially a node $\eta'$ is created that represents a star with predicate  $P_{\eta'} \doteq P_{\eta} \wedge P^j$. Informally, this step corresponds to assigning valuation $1$ to the decision variable $z_j$. There are $4$ scenarios that are evaluated. (i) If predicate $P_{\eta'}$ is infeasible, terminal $t_0$ is assigned as the $high$ child of $\eta$ (line~\ref{alg2:t0-as-high}). (ii) If $\eta'$ is at the terminal level ($k$) and $P_{\eta'}$ is feasible, $t_1$ is assigned as the $high$ child of $\eta$ (line~\ref{alg2:t1-as-high}). The corresponding path (beginning from root) indicates a valid characterization. 
Otherwise, (iii) either an isomorphic node is identified in $\mathcal{V'}$ and assigned as the  $high$ child (lines~\ref{alg2:iso-start}-~\ref{alg2:iso-end}) of $\eta$, or (iv) $\eta'$ is assigned as the $high$ child of $\eta$ and added to the set $\mathcal{V'}$ (lines~\ref{alg2:ubar-as-high} and~\ref{alg2:update-vprime}). Same steps are performed for computing the $low$ child of node $\eta$ by creating a node $\bar{\eta}'$ which denotes a star with predicate  $\bar{P}_{\eta'} \doteq P_{\eta} \wedge \bar{P}^j$. The \algnm{\algfnt{isomorphs}}  algorithm is discussed in the next section on BDD reduction. 

\algnm{\algfnt{3. traverse}}: It is a recursive routine which traverses $\mathcal{G}$ in a depth-first manner and maintains a regular expression $rex$ to track the nodes on each path in $\mathcal{G}$. A path terminating at $t_1$ indicates a unique characterization of the safety violation hence it is added to $\mathcal{C}_{\unsafe}$. Therefore, the final set $\mathcal{C}_{\unsafe}$  returned by the top level  \algnm{\algfnt{traverse}} as called in Algorithm~\ref{alg:construct-bdd} provides complete  characterization of counterexamples.

\begin{theorem}[Correctness]
\label{thm:correctness}
The set $\mathcal{C}_{\unsafe}$ returned by  \algnm{\algfnt{construct\_bdd}} algorithm (without reduction) is the complete characterization of counterexamples of length $k \in \mathbb{N}$ to a given safety specification $\phi_I^{\Psi}$.
\end{theorem}

\begin{proof}
We note that $C_{\unsafe}$ is the set of all binary strings of length $k$ accepted by the BDD. We denote $C_{\unsafe}^j$ to be the set of length $j \leq k$ prefixes of the elements in $C_{\unsafe}$, so we have $C_{\unsafe} = C_{\unsafe}^k$. 
We next define $R$ which is a set of all feasible binary strings of length $k$.
$$
R \deq \{ (z_1, z_2, \ldots z_k) \: |\: \bigwedge_{i=1}^k Q_i (z_i) (\bar{\alpha}) = \top, z_i \in \mathbb{B} \}, \mbox{ where }
$$
$$
Q_i (z_i) = 
\begin{cases}
    P^i ,& \text{if } z_i = 1\\
    \bar{P}^i, & \text{if } z_i = 0.
\end{cases}
$$

\noindent{It suffices to prove that $C_{\unsafe} = R$ in order to show that all strings accepted by the BDD are indeed feasible and all strings that are not accepted are not valid counterexamples.}

We first show that $R \subseteq C_{\unsafe}$. Consider an element $z \doteq (z_i, z_2, \ldots, z_k) \in R$.  Assuming that $z \notin C_{\unsafe}$, there exists a prefix $(z_1, \ldots, z_j) \in C_{\unsafe}^j$ of $z$, where $j = \mbox{min} \{ j' \in [1,k] \: | \: (z_1, \ldots, z_{j'}) \in C_{\unsafe}^{j'} \mbox{ and } (z_1, \ldots, z_{j'+1}) \notin C_{\unsafe}^{j'+1}\}$. Informally, $(z_1, \ldots, z_{j'+1})$ denotes the \emph{minimal infeasible prefix} of $z$. It also follows from the BDD construction that $(z_1, \ldots, z_{j'+1}) \notin C_{\unsafe}^{j'+1}$ \emph{iff} $\bigwedge_{i=1}^{j'+1} Q_i (z_i) (\bar{\alpha}) = \bot$. This further implies that $\bigwedge_{i=1}^{k} Q_i (z_i) (\bar{\alpha}) = \bot$ (i.e., $z \notin R$), which is a contradiction because $b \in R$. Therefore, all feasible strings of length $k$ are accepted by the BDD.

We next show that $C_{\unsafe} \subseteq R$. Consider an element $b \doteq (z_i, z_2, \ldots, z_k) \in C_{\unsafe}$. Now, assuming that $z \notin R$, we have  $\bigwedge_{i=1}^k Q_i (z_i)(\bar{\alpha}) = \bot$ (by definition of $R$). This implies that $\exists j \in [1, k]$ such that $\bigwedge_{i=1}^j Q_i (z_i)(\bar{\alpha}) = \bot$ and  $\bigwedge_{i=1}^{j-1} Q_i (z_i)(\bar{\alpha}) = \top$. It follows that $(z_1, \ldots, z_{j}) \notin C_{\unsafe}^{j}$. And since $(z_1, \ldots, z_{j})$ is a prefix of $z$, and is infeasible, we have $z \notin C_{\unsafe}$, which is a contradiction. Therefore, all strings captured by the BDD are indeed feasible.
\end{proof}

\begin{algorithm}
\SetKwInOut{Input}{input}\SetKwInOut{Output}{output}\SetKw{Return}{return}
\Input{node: $\eta$, level: $j$, set of nodes at level $j$: $\mathcal{V}'$, ordered sets $\Theta_{\Pi}^{\unsafe}$ and $\Theta_{\Pi}^{\neg \unsafe}$}
\Output{updated set of nodes at level $j$: $\mathcal{V}'$}
$k \gets |\Theta_{\Pi}^{\unsafe}|$\tcp*{\# of decision variables}
$\eta' \gets \algnm{\algfnt{create\_node}}(\Star_{\eta} \cap \Theta_{\Pi}^{\unsafe}[j])$\tcp*{for $z_j = 1$}
\If{$(P_{\eta'} (\bar{\alpha})$ == $\bot)$}{$\eta.high \gets \mathcal{G}.t_0$\label{alg2:t0-as-high}\tcp*{\emph{high} child of $\eta$ is $t_0$}
}
\ElseIf{$( j == k)$}{
{$\eta.high \gets \mathcal{G}.t_1$\label{alg2:t1-as-high}\tcp*{a valid characterization}
}
}
\Else{
    $Q^{\unsafe} \gets \Theta_\Pi^{\unsafe} [ j+1 \rightarrow k]$\;
    $Q^{\neg \unsafe} \gets \Theta_{\Pi}^{\neg \unsafe} [ j+1 \rightarrow k]$\;
    $\mathcal{J} \gets Q^{\unsafe} \cup Q^{\neg \unsafe}$\;
    \For{$\eta'' \in \mathcal{V}'$}{\label{alg2:iso-start}
        \If{\textnormal{\algnm{\algfnt{isomorphs}}}$\left(\eta', \eta'', \mathcal{J}\right)$ == $\top$}{
        $\eta.high \gets \eta''$\label{alg2:isomorphs}\tcp*{$\eta'$, $\eta''$ are isomorphs}
        }
    \label{alg2:iso-end}}
    \If{$\eta.high == \bot$}{
    $\mathcal{G}$.\algnm{\algfnt{add\_node}}($\eta'$)\tcp*{if no iso. node found}
    $\eta.high \gets \eta'$\;\label{alg2:ubar-as-high}
    $\mathcal{V}' \gets \mathcal{V}' \cup \eta'$\label{alg2:update-vprime}\tcp*{add $\eta'$ to $\mathcal{V'}$}
    }
}
$\bar{\eta}' \gets  \algnm{\algfnt{create\_node}}(\Star_{\eta} \cap \Theta_{\Pi}^{\neg \unsafe}[j])$\tcp*{for $z_j = 0$}
repeat steps 3-23 with $\bar{\eta}'$ for adding $low$ child at $\eta$\;
{\bf return} $\mathcal{V}'$\;
\caption{\small \algnm{\algfnt{process\_node}} algorithm. The algorithm adds $low$ and $high$ children of the input node $\eta$. Additionally, it  performs \emph{isomorphism} to identify an existing node that can serve as a child of $\eta$, to potentially keeping a check on the BDD size.}
\label{alg:process-node}
\end{algorithm}

\begin{algorithm}
\SetKwInOut{Input}{input}\SetKwInOut{Output}{output}\SetKw{Return}{return}
\Input{BDD node: $\eta$, regular expression: $rex$, character $ch$, set of unique paths: $\mathcal{C}_{\unsafe}$}
\Output{updated set $\mathcal{C}_{\unsafe}$}
$rex \gets rex + ch$\tcp*{string concatenation}
\If{$\eta == \mathcal{G}.t_1$}{
    $\mathcal{C}_{\unsafe} \gets \mathcal{C}_{\unsafe} \cup rex$\tcp*{$rex$ is a valid characterization}
}
\ElseIf{$\eta \neq \mathcal{G}.t_0$}{
    $\mathcal{C}_{\unsafe} \gets \algnm{\algfnt{traverse}}(\eta.low, rex, `0', \mathcal{C}_{\unsafe}$)\;
    $\mathcal{C}_{\unsafe} \gets \algnm{\algfnt{traverse}}(\eta.high, rex, `1', \mathcal{C}_{\unsafe}$)\;
}
{\bf return} $\mathcal{C}_{\unsafe}$\;
\caption{BDD \algnm{\algfnt{traverse}} algorithm.}
\label{alg:traverse-bdd}
\end{algorithm}

\subsection{Feasibility Model}
The feasibility of predicate $P_{\eta'}(\bar{\alpha}) \deq A \bar{\alpha} \leq b$ for the node $\eta'$ in  \algnm{\algfnt{process\_node}} algorithm is examined by solving following optimization problem. 
\begin{equation*}
\begin{aligned}
& \max ~d  \\
& \textnormal{s.t. } (\state{a}_{j})^T \bar{\alpha} \leq b_j,  1 \leq j \leq \lvert P_{\eta'} \rvert,
\end{aligned}
\end{equation*}
where $\lvert P_{\eta'} \rvert$ is the number of constraints in predicate $P_{\eta'}$. Here, $d \in \reals$ is a constant as certain optimization models require explicit specification of the objective function. This problem can be modeled as either a Linear Program (LP) or a  Satisfiability Modulo Theories (SMT) query. A satisfiable valuation of basis variables $\bar{\alpha}$ is an evidence to $P_{\eta'}$ feasibility. The feasibility of predicate $\bar{P}_{\eta'}$ is examined in a similar manner. The satisfiable valuation of $\bar{\alpha}$ obtained at terminal $t_1$ (line~\ref{alg2:t1-as-high}) gives a representative counterexample for corresponding characterization. 

\begin{figure*}[ht]
\centering     
\quad \quad \subfigure[The output of \algnm{\algfnt{create\_bdd}} algorithm. The solid edge denotes assignment $1$ to corresponding decision variable while dashed edge represents assignment $0$. The diagram represents all $10$ characterizations of the safety violation. The total number of nodes ($\mathcal{N}$) are $21$ (including $t_0$) and the width ($\mathcal{W}$) is $7$.]{\label{fig:default-bdd-basic}\includegraphics[width=0.42\linewidth,height=70mm]{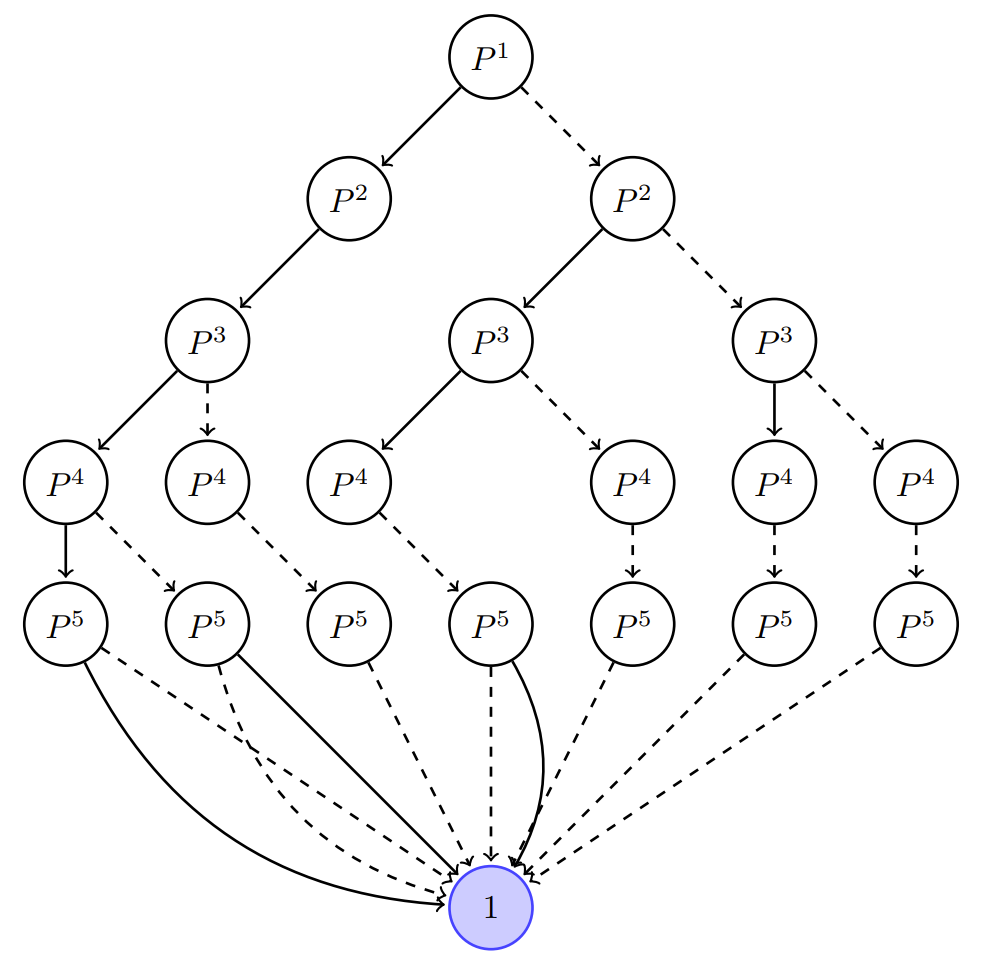}} \hfill
\subfigure[Multiple characterizations of the given safety violation and their representative counterexamples. The counterexample $11100$ represents the characterization where the reachable set violates the specification at $3^{rd}, 4^{th}, 5^{th}$ time steps; whereas the execution $00100$ denotes the characterization where the reachable set violates the specification at only $5^{th}$ time step.] {\label{fig:osc-counterexamples}\includegraphics[width=90mm, height=57mm]{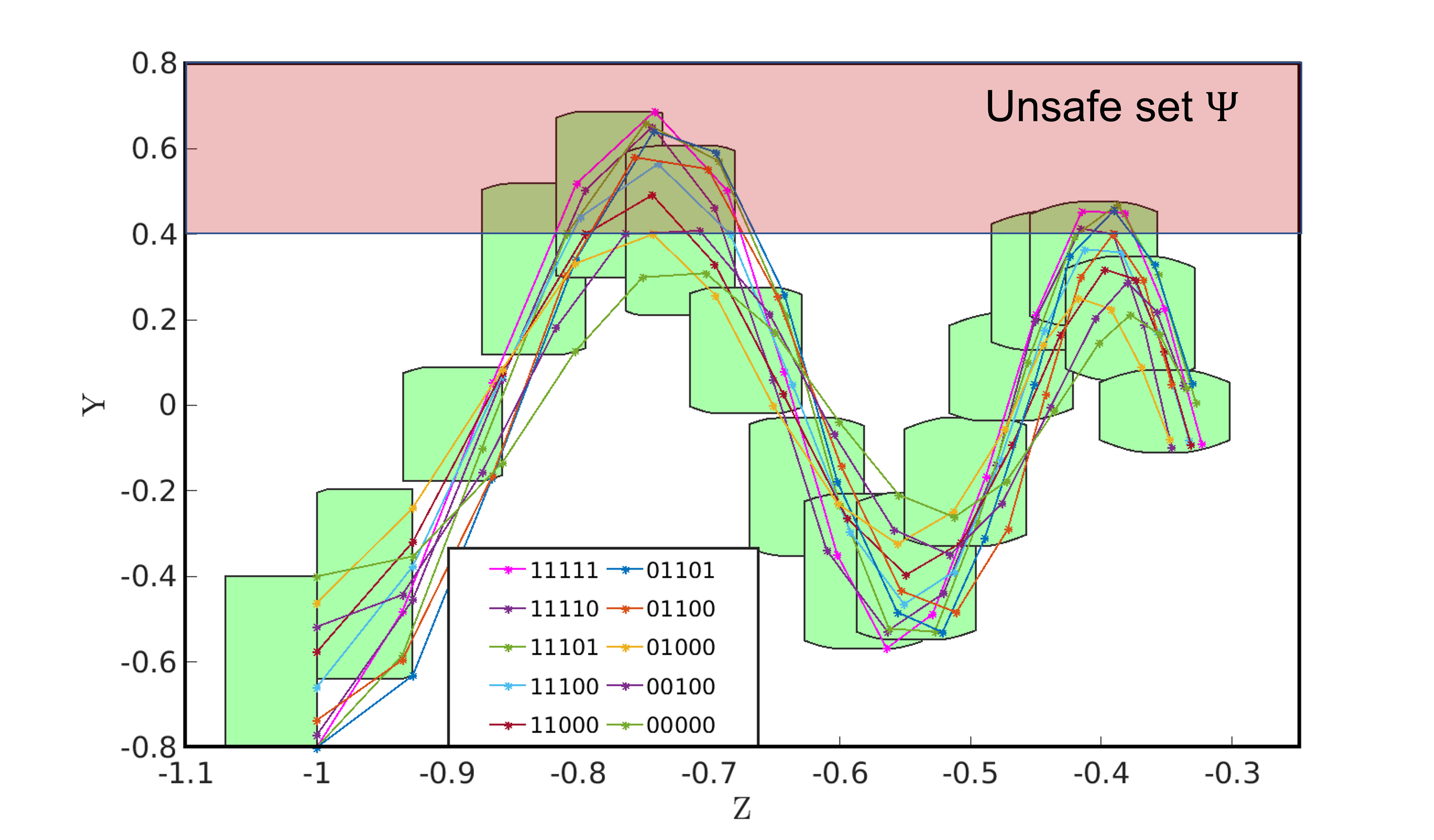}}
\caption{All valid characterizations of the safety violation}
\label{fig:characterization}
\end{figure*}

\textbf{Demonstration:} The BDD constructed for Example~\ref{example:osc-particle} with ordering [1, 2, 3, 4, 5] is shown in Fig.~\ref{fig:default-bdd-basic}. At each variable node, the \emph{solid} edge denotes $high$ child while \emph{dashed} edge is for the $low$ child. 
We omit terminal node $t_0$ and all of its incoming edges (due to the in-feasibility of predicate $P_{\eta'}$) for clarity. For example, the $low$ child of the left $P^2$ node, which is not shown, is $t_0$ because $(P^1 \wedge \bar{P}^2) (\bar{\alpha}) = \bot$. These feasibility results can also be easily verified with the help of Fig.~\ref{fig:theta-constr-prop}.
The decision diagram has total ($\mathcal{N}$) $21$ nodes (including $t_0$) and has width ($\mathcal{W}$) $7$. The number of unique paths from \emph{root} to  terminal $t_1$ is $10$ which is also the number of characterizations ($\mathcal{C}_{\unsafe}$) of the safety violation under consideration. Some of these paths are $11110, 11101, 11000, 01100$, and $00100$. The system executions generated from respective valuations of $\bar{\alpha}$ are illustrated in Fig.~\ref{fig:osc-counterexamples}.

\textbf{Complexity Analysis:} In the absence of conducting isomorphism among nodes (lines 9-22) in \algnm{\algfnt{process\_node}} algorithm, its run time complexity is directly proportional to the complexity of checking the feasibility of node $\eta'$. The complexity of running feasibility check on a node is polynomial in the total number of its variables which includes continuous basis variables as well as binary decision variables. Note that in an $n$- dimensional system with $m$ bounded inputs, $m$ input basis variables and $2m$ constraints are added to  each successive star. If $0 < j < \timebound$ is the largest time instance at which the reachable set has an non-empty overlap with the unsafe set $\Psi$, then the total number of basis variables are $n + j \times m$. Further, if the reachable set overlaps with $\Psi$ at $k$ different time instances, we check the feasibility of $k$ stars in the worst case (i.e., for a leaf node). Since the number of constraints in the feasibility model is linear in the number of variables, the running time of checking the feasibility of a node is $\mathcal{O}((n+m)k')$ where $k' = \beta k$ for some $\beta \leq \timebound$. However, the number of feasibility instances we solve can be exponential in $k$ which the number of time instances at which the reachable set has an non-empty overlap with the unsafe set. Thus the running time of BDD construction algorithm \algnm{\algfnt{construct\_bdd}} is exponential in $k$.

Due to the number of feasibility instances we solve, the size of the BDD  can grow exponentially large in $k$ in the worst case.  The number of feasibility instances is highly dependent on the order of decision variables. For our small running example, a random ordering of the variables results into the diagram with $23$ nodes and width $8$ (compared to $21$ nodes and width $7$ for the default ordering). The size also provides an estimate on the number of times that one needs to solve feasibility of a sub-characterization. There can also be isomorphic nodes at each level and applying the \emph{isomorphism} rule can significantly reduce the number of feasibility instances to be solved as well as size of the diagram.
\section{BDD Reduction}

In this section, we discuss a technique to obtain a reduced BDD by applying isomorphism to its nodes. We do not add a new node to the partially constructed decision diagram if we discover an existing BDD node to be isomorphic to this new node. We note that BDD nodes are generalized stars with the same center and the basis vectors but possibly different predicates. 
We also note from Section~\ref{subsec:bdd} that two variable nodes are called \emph{isomorphic} to each other if they are labeled by the same decision variable (i.e., they are at the same level) and have the same set of paths starting from them. 
%
A path starting at a node is a binary string denoted as a sequence of \emph{high} and \emph{low} edges corresponding to a valid assignment to the remaining predicates in conjunction with the predicate of the current node. Consequently, two nodes having same set of paths originating at them would mean that both nodes have the same set of valid assignments to the remaining predicates.
Thus the problem of determining isomorphism among two nodes is same as determining \emph{equivalence} of their  predicates with respect to the remaining predicates. 

%


\subsection{Predicates' Equivalence}
We note from Remark~\ref{rem:covex-predicate} that assigning $1$ or $0$ to a remaining predicate $P^j$ corresponds to the selection of either $P^j$ or $\bar{P}^j$ where both these predicates represent half spaces separated by only one hyperplane. An unsafe set defined over multiple constraints would have yielded disjunction in $\bar{P}$ in which case the  presented approach would not be applicable. 

\begin{definition}[Equivalence of predicates]
\label{def:pred-equ}
Given two predicates $P'_{\eta}$ and $P''_{\eta}$, and a set of additional $2q$ predicates: $\mathcal{J} \deq \{P^1, \bar{P}^1, P^2, \bar{P}^2, \ldots, P^{q}, \bar{P}^{q}\}$, the predicates $P'_{\eta}$ and $P''_{\eta}$ are defined to be \emph{equivalent with respect to} $\mathcal{J}$ (denoted as $P'_{\eta} \equiv_{\mathcal{J}} P''_{\eta}$) \emph{iff} $\forall J \subseteq \mathcal{J}$ s.t. $J \neq \emptyset$ and $J(\bar{\alpha}) = \top$, either of the following holds:

\begin{itemize}
    \item $(P'_{\eta}\cap J)(\bar{\alpha}) = \top$ and $(P''_{\eta} \cap J)(\bar{\alpha}) = \top$
    \item $(P'_{\eta}\cap J)(\bar{\alpha}) = \bot$ and $(P''_{\eta} \cap J)(\bar{\alpha}) = \bot$.
\end{itemize}

Note that $J$ can not have both a predicate $P$ and its complement $\bar{P}$ because in that scenario the second case would vacuously hold due to $J(\bar{\alpha}) = \bot$. 
%

\begin{theorem}
The respective predicates $P'_{\eta}$ and $P''_{\eta}$ of two isomorphic nodes $\eta'$ and $\eta''$ satisfy Definition~\ref{def:pred-equ}.
\end{theorem}
\begin{proof}
The proof consists of two parts. First part proves that any two isomorphic nodes in the BDD would satisfy Definition~\ref{def:pred-equ}. Second part proves that if two nodes satisfy Definition~\ref{def:pred-equ}, then they are isomorphic.

\noindent \textbf{Part 1:} We prove this using a proof by contradiction.
Assume for two isomorphic nodes $\eta'$ and $\eta''$, $\exists J \subseteq \mathcal{J}$, $J \neq \emptyset$, $J(\bar{\alpha}) = \top$ s.t. $(P'_{\eta}\cap J)(\bar{\alpha}) = \top$ and $(P''_{\eta}\cap J)(\bar{\alpha}) = \bot$. 
Let $I = \{i_1, i_2, \ldots, i_l\}$ be indices such that either $P^{i_m}$ or $\bar{P}^{i_m}$ is in $J$ and let $\bar{I} = \{1, 2, \ldots, q\} \setminus I$.
Consider the partial path $\rho$ that assigns high edge for the variable $i_m$ if $P^{i_m} \in J$ and assigns a low edge for $i_m$ if $\bar{P}^{i_m} \in J$.
If $P'_{\eta} \cap J = \top$ and $P''_{\eta} \cap J = \bot$, then the partial path assigned in $\rho$ is feasible for $P'_{\eta}$ and infeasible for $P''_{\eta}$.
Therefore, if one considers all possible $2^{q-l}$ paths obtained by assigning high and low edges to indices in $\bar{I}$, there exists at least one feasible path for $P'_{\eta}$, whereas, all such $2^{q-l}$ paths would be infeasible for $P''_{\eta}$.
Therefore, there exists at least one path that differentiates the predicates $P'_{\eta}$ and $P''_{\eta}$, a contradiction since we assume that $\eta'$ and $\eta''$ are isomorphic.

\noindent \textbf{Part 2:} Suppose that there are two nodes $\eta'$ and $\eta''$ that satisfy Definition~\ref{def:pred-equ}.
Consider a feasible path $\rho$ of high and low edges over indices $\{1, 2, \ldots, q\}$ for the node $\eta'$.
Consider the set $J_{\rho}$ to include predicates $P^{l}$ if the index $l$ is a high edge and $\bar{P}^l$ if the index $l$ is a low edge.
From the construction, $J_{\rho} \subseteq \mathcal{J}$, hence from Definition~\ref{def:pred-equ}, $P''_{\eta} \cap J_{\rho}$ is also feasible.
Therefore, the path $\rho$ is also feasible for $\eta''$.
That is, all feasible paths for $\eta'$ are also feasible for $\eta''$.
Similarly, all infeasible paths for $\eta'$ are also infeasible for $\eta''$.
Therefore, the nodes $\eta'$ and $\eta''$ are isomorphic.
%
\end{proof}

Definition~\ref{def:pred-equ} can alternatively be written as $P'_{\eta} \equiv_{\mathcal{J}} P''_{\eta}$ \emph{iff} $\nexists J \subseteq \mathcal{J}$, $J \neq \emptyset$, for which either of the following holds:

\begin{itemize}
    \item $(P'_{\eta}\cap J)(\bar{\alpha}) = \top$ and $(P''_{\eta} \cap J)(\bar{\alpha}) = \bot$
    \item $(P''_{\eta} \cap J)(\bar{\alpha}) = \top$ and $(P'_{\eta}\cap J)(\bar{\alpha}) = \bot$.
\end{itemize}

Informally, if there is a feasible assignment to the subset of predicates in $\mathcal{J}$ which distinguishes $P'_{\eta}$ and $P''_{\eta}$ then it is an evidence that $P'_{\eta} \not \equiv P''_{\eta}$. Thus maximizing over the cardinality of $J$ is sufficient to prove  the existence of such a distinguishable assignment. This brings us to the following problem statement.
\end{definition}

\textbf{Problem Statement II :} Given two predicates $P'_{\eta}$ and $P''_{\eta}$, and a set $\mathcal{J} \deq \{P^j, \bar{P}^j\}, 1 \leq j \leq q$ , find $J_{max} \doteq J'_{max} \cup J''_{max}$ where 

\begin{itemize}
    \item $J'_{max} \deq max_{|J|} \{J \subseteq \mathcal{J} \textnormal{ s.t. } (P'_{\eta}\cap J)(\bar{\alpha}) = \top$ and $(P''_{\eta} \cap J)(\bar{\alpha}) = \bot \}$
    \item $J''_{max} \deq max_{|J|} \{J \subseteq \mathcal{J} \textnormal{ s.t. } (P''_{\eta}\cap J)(\bar{\alpha}) = \top$ and $(P'_{\eta} \cap J)(\bar{\alpha}) = \bot \}$.
\end{itemize}

It follows that the equivalence of two predicates $P'_{\eta} \deq A'_{1} \bar{\alpha} \leq \state{b'}_1$ and $P''_{\eta} \deq A''_{1} \bar{\alpha} \leq \state{b''}_1$ with respect to the set of future predicates, $\mathcal{J}$,  can be performed in two analogous phases. First phase would be to find $J'_{max}$ to distinguish  $P'_\eta$ from $P''_{\eta}$. If $J'_{max} = \emptyset$, we attempt to find $J''_{max}$ in the second phase. If $J'_{max} = J''_{max} = \emptyset$, then $P'_{\eta} \equiv_{\mathcal{J}} P''_{\eta}$.

To enumerate over all possible distinguishable assignments $J$ and obtaining a largest among them, we use a mathematical tool called \emph{Farkas' Lemma} which is a theorem of alternatives for a finite set of linear constraints. It has several variants among which we adopt the one given below.

\begin{lemma}[Farkas' Lemma]
Let $A \in \reals^{r \times s}$ and $\state{b} \in \reals^s$. Then exactly one of the following two assertions holds~\cite{2006_Farkas}:
\begin{enumerate}
    \item $\exists \bar{\alpha} \in \reals^s$ such that $A \bar{\alpha} \leq \state{b}$;
    \item $\exists \state{y} \in \reals^r$ such that $A^{T} \state{y} = 0, \state{b}^T \state{y} < 0$, and $\state{y} \geq 0$.
\end{enumerate}
\end{lemma}


\subsection{Equivalence Formulation}

In this section, we explain the system of equations for checking the existence of $J'_{max}$; the equations for the existence check of $J''_{max}$ are same except that the roles of predicates $P'$ and $P''$ would be switched. 
Each system of equations consists of two sub-systems: one for the set of predicates $\{P'_{\eta}, \mathcal{J}\}$ while the other for the set $\{P''_{\eta}, \mathcal{J}\}$. 
Both sub-systems are linked together with the help of decision variables such that a feasible set of predicates, $J \subseteq \mathcal{J}$, in the first sub-system is forced to violate in the second sub-system using Farkas' lemma. Finally, an optimization problem over those decision variables is solved to find a largest $J$, denoted as $J'_{max}$. We now expand on this in a step-wise manner. 

We know from Definition~\ref{def:pred-equ} that  $\mathcal{J}$ is set of $2q$ convex predicates, thus we denote these elements as $A_j\bar{\alpha} \leq b_j$, $2 \leq j \leq 2q+1$ for the ease of discussion.

\textbf{Step 1}: The sub-system of predicates $\{P'_{\eta}, \mathcal{J}\}$ can be represented as follows

\begin{equation}
\label{eq:sys1-eq1} 
\begin{pmatrix}
A'_{1} \\
A_{2} \\
A_{3} \\
\vdots\\
A_{2q+1} 
\end{pmatrix}
\begin{pmatrix}
\bar{\alpha}
\end{pmatrix}\leq
\begin{pmatrix}
\state{b}'_{1} \\
\state{b}_{2} \\
\state{b}_{3} \\
\vdots\\
\state{b}_{2q+1}
\end{pmatrix}
\end{equation}

or, 
\begin{align}
H\bar{\alpha}\leq \state{g}.
\label{eq:sys1-hxg1}
\end{align}

\noindent{This sub-system~\ref{eq:sys1-hxg1} is \emph{feasible} if it has a satisfiable valuation of $\bar{\alpha} \in \mathbb{R}^n$. 
We denote $(H'_J\bar{\alpha} \leq \state{g'}_J) \doteq P'_{\eta} (\bar{\alpha}) \wedge \left(\bigwedge_{P^j \in J} P^j (\bar{\alpha})\right)$.} \\

\noindent{Similarly, the sub-system with predicates $\{P''_{\eta}, \mathcal{J}\}$ is given as}

\begin{equation}
\label{eq:sys2-eq1} 
\begin{pmatrix}
A''_{1} \\
A_{2} \\
A_{3} \\
\vdots\\
A_{2q+1} 
\end{pmatrix}
\begin{pmatrix}
\bar{\alpha}
\end{pmatrix}\leq
\begin{pmatrix}
\state{b''}_{1} \\
\state{b}_{2} \\
\state{b}_{3} \\
\vdots\\
\state{b}_{2q+1}
\end{pmatrix}
\end{equation}

or, 
\begin{align}
H''\bar{\alpha}\leq \state{g''}.
\label{eq:sys2-hxg1}
\end{align}

\noindent{This sub-system~\ref{eq:sys2-hxg1} is infeasible if $\exists \state{y''} \geq 0$ such that} 
\begin{align}
& H''^{T}\state{y''}=0, \state{g''}^{T}\state{y''} < 0. 
\label{eq:sys2-hxg1-farkas}
\end{align}

\textbf{Step 2}: Since number of variables in $\state{y''}$ is same as the number of constraints in  sub-system~\ref{eq:sys2-hxg1}, we have $\state{y''} = [y''_{1,1}, y''_{1,2}, \ldots, y''_{2,1}, y''_{2,2},\ldots, y''_{2q+1, |P^{2q+1}|}]^T$. Further, we use $y''_J$ to denote $\state{y''}$ variables that correspond to the set  $\{P'_{\eta}, J\}$. For example, $y''_J = [y''_{1,1}, \ldots, y''_{1,|P''_{\eta}|}, y''_{2,1},y''_{2,2}, \ldots, y''_{q, |P^q|}]^T$ for $J = \{P^2, \ldots, P^q\}$.
This is same as dropping the rest of $\state{y''}$ variables from system or assigning $0$ to them, i.e., $\state{y''}_J = [y''_{1,1}, \ldots, y''_{1,|P''_{\eta}|}, 0, \ldots, 0, y''_{2,1},y''_{2,2}, \ldots, y''_{q, |P^q|}, 0, \ldots, 0]^T$. This is essentially same as dropping columns related to predicates $P^{q+1}, P^{q+2}, \ldots, P^{2q+1}$ from matrix $H''^T$ in sub-system ~\ref{eq:sys2-hxg1-farkas}. As we are only interested in examining the feasibility w.r.t. the set $J$, this step ensures that the predicates in the set $\mathcal{J}\symbol{92} J$ are dropped  from both the sub-systems. The rationale is that we are only interested in find a set $J$ such that $P'_{\eta} \cap J$ is feasible and  $P''_{\eta} \cap J$ is infeasible so we should drop the predicates in the set $\mathcal{J}\symbol{92} J$ from the system.



Thus the problem of finding a set $J$ is reduced to checking feasibility of the following system.
\begin{align}
& (H'_J \bar{\alpha} \leq \state{g'}_J), & &  H''^{T}\state{y''}=\state{0}, & & \state{g''}^{T}\state{y''} < 0, &  \state{y''}_J > 0.
\label{eq:case1-final-solution}
\end{align}

%

\noindent{Finally, we run an optimization solver on this system to find $J'_{max}$.
Informally, a non-empty subset $J'_{max}$ indicates an assignment to the future decision variables that can make a path between $P'_{\eta}$ and terminal $t_1$ distinguishable from any path between $P''_{\eta}$ and $t_1$. On the other hand, $J'_{max} = \emptyset$  would mean the non-existence of such a distinguishable path,  which would imply that the set of valid paths starting at $P'_{\eta}$ is contained in the set of valid paths starting at $P''_{\eta}$.

Recall that the second phase to checking the equivalence of $P'_{\eta}$ and $P''_{\eta}$ entails that the set of valid paths starting at $P'_{\eta}$ is contained in the set of valid paths starting at $P''_{\eta}$ for which we simply swap the roles of two sub-systems formulated in the first phase.}

%
%
%


\noindent{
\textbf {Demonstration:} Consider that $P'_{\eta} \doteq P^1 \wedge  P^2 \wedge \bar{P}^3$, $P''_{\eta} \doteq P^1 \wedge P^2 \wedge P^3$, and $\mathcal{J} = \{P^4, \bar{P}^4, P^5, \bar{P}^5\}$ in Example~\ref{example:osc-particle} (The variables ordering  is [1, 2, 3, 4, 5]). The solution to system~\ref{eq:case1-final-solution} with these parameters would be $J'_{max} = \emptyset$, which means that all valid paths starting at $P'_{\eta}$ are also valid for $P''_{\eta}$. However, the system with these predicates swapped i.e., $P'_{\eta} \doteq  P^1 \wedge P^2 \wedge P^3$, $P''_{\eta} \doteq P^1 \wedge P^2 \wedge \bar{P}^3$ would yield any of the 3 solutions (depending on the optimization solver): $\{P^4,P^5\},  \{\bar{P}^4, P^5\}, \{P^4, \bar{P}^5\}$. It follows that there is a valid path at node $P'_{\eta}$ that differs from those at node $P''_{\eta}$, %
hence these predicates $P'_{\eta}$ and $P''_{\eta}$ are \emph{not equivalent w.r.t.} the remaining predicates. It can also be verified using Fig.~\ref{fig:default-bdd-basic} that the node with predicate $P^1 \wedge  P^2 \wedge P^3$ has $4$ valid paths - $10, 11, 00, 01$; whereas the node with predicate $P^1 \wedge  P^2 \wedge \bar{P}^3$ has $00$ as its only valid path. }

\textbf{Complexity analysis:} Each new node $\eta'$ at a level is checked for its isomorphic counterpart among all the existing node at that level with respect to the nodes at remaining (future) levels or all possible future paths. At any given (current or future) level $j$, the number of nodes can be exponential in $j$. Since there are $k$ levels in the decision diagram, the run-time complexity of constructing the reduced BDD is still exponential in $k$.

\begin{figure}
\centering     
\includegraphics[width=60mm]{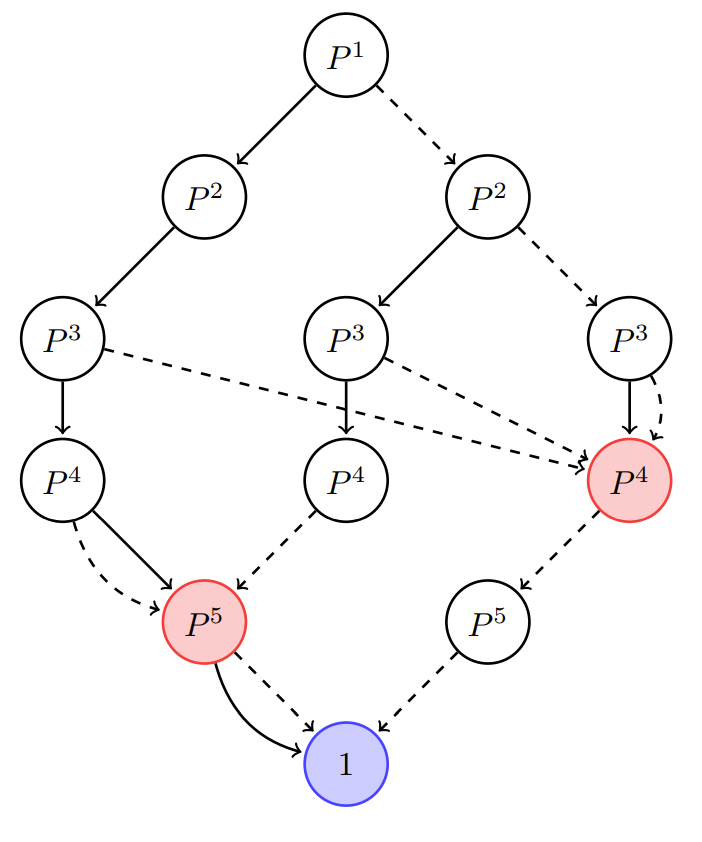}
\caption{\small The reduced decision diagram obtained upon performing \emph{isomorphism} among variable nodes at each level. The red-colored nodes are the ones identified as \emph{isomorphs} to some other node(s) during the BDD construction.}
\label{fig:reduced-bdd}
\end{figure}

\subsection{MILP Model for Equivalence}
We implement system~\ref{eq:case1-final-solution} as a Mixed Integer Linear Program and feed it to an optimization solver. We introduce $2q+1$ binary variables $z'_1, z'_2, \ldots, z'_{2q+1}$ for $2q+1$ predicates in system~\ref{eq:sys1-hxg1}. Then the model (with sufficiently large $M$), $\forall i \in [1, 2q+1]$, becomes
\begin{align}
& \max \sum_i z'_i \nonumber\\
& \textnormal{s.t. } \left(\state{a'}_{i,j}\right)^T \bar{\alpha} \leq b'_{i, j} + M \left( 1 - z'_i \right), \label{eq:model1-encode-eq1}\\
& z'_i \in \{0, 1\}, \bar{\alpha} \in \mathbb{R}^{n}, j = 1, \ldots, |P_i| \nonumber.
\end{align}

\noindent{\textbf{Note:} The variables $z'_i$ that are assigned $1$ by the optimization solver denote predicates in $J'$. As we are interested in finding $J'_{max} \subseteq \mathcal{J}$, we maximize over $z'_i$ variables. An alternative is a constant objective function with additional constraints explicitly specifying that at least one of the $z'_i$ variables is assigned $1$ by the solver.}

\noindent{Next, the following set of constraints encodes system~\ref{eq:sys2-hxg1-farkas}.}
\begin{align}
& \left(H''\right)^T\state{y''}\leq 0, & & \left(H''\right)^T\state{y''}\geq 0, & \state{g''}^T \state{y''} \leq \epsilon\label{eq:model1-encode-hxg2} \nonumber\\
& \textnormal{s.t. } y''_{i, j} \geq 0, & & 0 < \epsilon \ll 1.
\end{align}

\noindent{Observe that $z'_i$ variables assigned $0$ by the solver in model~\ref{eq:model1-encode-eq1} are associated with the set $\mathcal{J}\symbol{92}J$. The corresponding $y''_{i}$ variables in model~\ref{eq:model1-encode-hxg2} are required to be assigned $0$ as well. The next set of constraints enforces this requirement and completes the encoding for system~\ref{eq:case1-final-solution} by combining models~\ref{eq:model1-encode-eq1} and~\ref{eq:model1-encode-hxg2}.}
\begin{align}
    y''_{i, j} \leq M * z'_i, \nonumber \\
    y''_{i, j} \geq \epsilon * z'_i.
\label{eq:model1-encode-final-sol1} \end{align}

The model for $J''_{max}$ would be similar. The reduced binary decision diagram for the running example is shown in Fig.~\ref{fig:reduced-bdd}. As compared to the original BDD, the total number of nodes $\mathcal{N}$ is reduced from $21$ to $13$ (including $t_0$) and the width $\mathcal{W}$ is reduced from $7$ to $3$.

\section{Evaluation and Discussion}
\begin{table*}
\scriptsize
\centering
\begin{tabular}{| *{14}{c|} }
\hline
& & & & & & & & \multicolumn{3}{c|}{} & \multicolumn{3}{c|}{} \\
& & & Bounded & $\unsafe$-overlap & Verification & Decision & &  \multicolumn{3}{c|}{OBDD} & \multicolumn{3}{c|}{ROBDD} \\
\cline{9-14}
~S. No.~ & System & Dims & Inputs & interval(s) & Time & variables & $|\mathcal{C}_{\unsafe}|$ & & & & & & \\
& & ($n$) & ($m$) &  & (sec) & ($k$) & & $\mathcal{N}$ & $\mathcal{W}$ & $\mathrm{T_c}$ & $\mathcal{N}$ & $\mathcal{W}$ & $\mathrm{T_c}$ \\
& & & & & & & & & & (sec) & & & (sec)  \\
\hline \multirow{3}{*}{\#1} & \multirow{3}{*}{Osc Particle} & \multirow{3}{*}{3} & \multirow{3}{*}{1} & \multirow{3}{*}{[3,5][12,13]} & \multirow{3}{*}{1} & \multirow{3}{*}{5} & \multirow{3}{*}{10} & 21 & \multirow{2}{*}{7} & \multirow{3}{*}{2} & 13 & \multirow{2}{*}{3} & 6 
  \\
 \cdashline{9-9}
 \cdashline{12-12}
 \cdashline{14-14}
 & & & & & & & & 22 & &  & 13 & & 7 \\
  \cdashline{9-10}
  \cdashline{12-14}
  & & & & & & & & 23 & 8 & & 14 & 4 & 5 \\
 \hline \multirow{3}{*}{\#2} & \multirow{3}{*}{Osc Particle} & \multirow{3}{*}{3} & \multirow{3}{*}{2} & \multirow{3}{*}{[3,5][12,13]} & \multirow{3}{*}{2} & \multirow{3}{*}{6} & \multirow{3}{*}{17} & 39 & \multirow{3}{*}{16} & 13 & 17 & \multirow{3}{*}{4} & 36 
  \\
 \cdashline{9-9}
 \cdashline{11-12}
 \cdashline{14-14}
 & & & & & & & & 40 & & 13 & 18 & & 38 \\
  \cdashline{9-9}
  \cdashline{11-12}
  \cdashline{14-14}
  & & & & & & & & 47 & & 15 & 17 & & 30\\
\hline \multirow{3}{*}{\#3} & \multirow{3}{*}{RS-14} & \multirow{3}{*}{3} & \multirow{3}{*}{1} & \multirow{3}{*}{[6, 21]} & \multirow{3}{*}{2} & \multirow{3}{*}{16} & \multirow{3}{*}{23} & 171 & \multirow{2}{*}{22} & 91 & 69 & 7 & 373
  \\
  \cdashline{9-9}
 \cdashline{11-14}
  & & & & & & & & 177 & & 95 & 55 & 5 & 256\\
    \cdashline{9-14}
  & & & & & & & & 166 & 21 & 77 & 95 & 10 & 487\\
  \hline \multirow{3}{*}{\#4} & \multirow{3}{*}{RS-14} & \multirow{3}{*}{3} & \multirow{3}{*}{2} & \multirow{3}{*}{[6, 22]} & \multirow{3}{*}{4} & \multirow{3}{*}{17} & \multirow{3}{*}{26} & 206 & \multirow{2}{*}{25} & 381 & 71 & 7 & 1531
  \\
 \cdashline{9-9}
 \cdashline{11-14}
 & & & & & & & & 214 & & 354 & 59 & 5 & 1117 \\
  \cdashline{9-14}
  & & & & & & & & 180 & 24 & 370 & 99 & 10 & 1610 \\
      \hline \multirow{3}{*}{\#5} & \multirow{3}{*}{RS-3} & \multirow{3}{*}{4} & \multirow{3}{*}{1} & \multirow{3}{*}{[19, 36]} & \multirow{3}{*}{3} & \multirow{3}{*}{18} & \multirow{3}{*}{27} & 212 & \multirow{2}{*}{26} & 388 & 92 & 9 & 1789  \\
 \cdashline{9-9}
 \cdashline{11-14}
 & & & & & & & & 224 & & 402 & 68 & 6 & 1185  \\
  \cdashline{9-14}
  & & & & & & & & 210 & 25 & 392 & 108 & 9 & 1662 \\
    \hline \multirow{3}{*}{\#6} & \multirow{3}{*}{RS-3} & \multirow{3}{*}{4} & \multirow{3}{*}{2} & \multirow{3}{*}{[19, 37]} & \multirow{3}{*}{5} & \multirow{3}{*}{19} & \multirow{3}{*}{37} & 294 & \multirow{2}{*}{36} & 1910 & 99 & 9 & 5861 \\
 \cdashline{9-9}
 \cdashline{11-14}
 & & & & & & & & 321 & & 2206 & 77 & 7 & 6228 \\
  \cdashline{9-14}
  & & & & & & & & 260 & 34 & 1790 & 115 & 10 & 7186  \\
  \hline \multirow{3}{*}{\#7} & \multirow{3}{*}{RS-6} & \multirow{3}{*}{8} & \multirow{3}{*}{1} & \multirow{3}{*}{[8, 22]} & \multirow{3}{*}{3} & \multirow{3}{*}{15} & \multirow{3}{*}{28} & 190 & \multirow{3}{*}{27} & 166 & 62 & 7 & 534  \\
 \cdashline{9-9}
 \cdashline{11-14}
 & & & & & & & & 204 & & 161 & 52 & 5 & 442  \\
 \cdashline{9-9}
  \cdashline{11-14}
  & & & & & & & & 182 & & 154 & 78 & 9 & 633 \\
   \hline \multirow{3}{*}{\#8} & \multirow{3}{*}{RS-6} & \multirow{3}{*}{8} & \multirow{3}{*}{2} & \multirow{3}{*}{[8, 23]} & \multirow{3}{*}{6} & \multirow{3}{*}{16} & \multirow{3}{*}{45} & 329 & \multirow{2}{*}{44} & 862 & 75 & 9 & 3211 \\
 \cdashline{9-9}
 \cdashline{11-14}
 & & & & & & & & 354 & & 890 & 62 & 6 & 2158 \\
 \cdashline{9-14}
  & & & & & & & & 264 & 37 & 707 & 112 & 14 & 4380 \\
    \hline \multirow{3}{*}{\#9} & \multirow{3}{*}{RS-8} & \multirow{3}{*}{8} & \multirow{3}{*}{1} & \multirow{3}{*}{[13, 30]} & \multirow{3}{*}{4} & \multirow{3}{*}{18} & \multirow{3}{*}{33} & 246 & \multirow{2}{*}{31} & 350 & 98 & 9 & 1891 \\
 \cdashline{9-9}
 \cdashline{11-14}
 & & & & & & & & 269 & & 390 & 73 & 7 & 1160 \\
 \cdashline{9-14}
  & & & & & & & & 228 & 30 & 374 & 102 & 9 & 1985 \\
\hline \multirow{3}{*}{\#10} & \multirow{3}{*}{RS-8} & \multirow{3}{*}{8} & \multirow{3}{*}{2} & \multirow{3}{*}{[13, 31]} & \multirow{3}{*}{6} & \multirow{3}{*}{19} & \multirow{3}{*}{46} & 343 & \multirow{2}{*}{45} & 1182 & 107 & 10 & 5380 \\
 \cdashline{9-9}
 \cdashline{11-14}
 & & & & & & & & 382 & & 1282 & 88 & 8 & 4312 \\
 \cdashline{9-14}
  & & & & & & & & 302 & 42 & 1166 & 122 & 10 & 4153 \\
\hline \multirow{3}{*}{\#11} & \multirow{3}{*}{RS-7} & \multirow{3}{*}{12} & \multirow{3}{*}{1} & \multirow{3}{*}{[9, 20]} & \multirow{3}{*}{3} & \multirow{3}{*}{12} & \multirow{3}{*}{26} & 137 & \multirow{3}{*}{25} & 72 & 47 & 7 & 285 \\
 \cdashline{9-9}
  \cdashline{11-14}
 & & & & & & & & 146 &  & 77 & 39 & 5 & 200  \\
\cdashline{9-9}
  \cdashline{11-14}
  & & & & & & & & 133 &  & 72 & 62 & 9 & 340 \\
  \hline \multirow{3}{*}{\#12} & \multirow{3}{*}{RS-7} & \multirow{3}{*}{12} & \multirow{3}{*}{2} & \multirow{3}{*}{[9, 21]} & \multirow{3}{*}{5} & \multirow{3}{*}{13} & \multirow{3}{*}{40} & 238 & \multirow{2}{*}{39} & 358 & 56 & 8 & 1250 \\
 \cdashline{9-9}
  \cdashline{11-14}
 & & & & & & & & 256 &  & 382 & 52 & 7 & 1145  \\
\cdashline{9-14}
  & & & & & & & & 192 & 38 & 307 & 75 & 10 & 1280 \\
    \hline \multirow{3}{*}{\#13} & \multirow{3}{*}{RS-7} & \multirow{3}{*}{12} & \multirow{3}{*}{3} & \multirow{3}{*}{[8, 23]} & \multirow{3}{*}{7} & \multirow{3}{*}{16} & \multirow{3}{*}{73} & 527 & \multirow{2}{*}{72} & 2029 & 82 & 9 & 5345 \\
 \cdashline{9-9}
  \cdashline{11-14}
 & & & & & & & & 595 &  & 2223 & 74 & 8 & 4534 \\
\cdashline{9-14}
  & & & & & & & & 591 & 58 & 1626 & 138 & 18 & 9870 \\
     \hline \multirow{3}{*}{\#14} & \multirow{3}{*}{RS-9} & \multirow{3}{*}{16} & \multirow{3}{*}{-} & \multirow{3}{*}{[51, 65]} & \multirow{3}{*}{3} & \multirow{3}{*}{15} & \multirow{3}{*}{33} & 205 & \multirow{3}{*}{32} & 42 & 74 & 9 & 192 \\
  \cdashline{9-9}
 \cdashline{11-14}
 & & & & & & & & 232 & & 47 & 63 & 7 & 168 \\
 \cdashline{9-9}
  \cdashline{11-14}
& & & & & & & & 204 & & 44 & 88 & 10 & 225 \\
     \hline \multirow{3}{*}{\#15} & \multirow{3}{*}{RS-9} & \multirow{3}{*}{16} & \multirow{3}{*}{1} & \multirow{3}{*}{[50, 66]} & \multirow{3}{*}{5} & \multirow{3}{*}{17} & \multirow{3}{*}{41} & 266 & \multirow{2}{*}{39} & 1443 & 93 & 10 & 6330 \\
  \cdashline{9-9}
 \cdashline{11-14}
 & & & & & & & & 300 & & 1529 & 79 & 8 & 5338 \\
 \cdashline{9-14}
& & & & & & & & 257 & 37 & 1389 & 129 & 14 & 8312 \\
  \hline \multirow{3}{*}{\#16} & \multirow{3}{*}{RS-10} & \multirow{3}{*}{20} & \multirow{3}{*}{-} & \multirow{3}{*}{[30, 43]} & \multirow{3}{*}{3} & \multirow{3}{*}{14} & \multirow{3}{*}{23} & 145 & \multirow{2}{*}{22} & 41 & 65 & 8 & 201 \\
  \cdashline{9-9}
 \cdashline{11-14}
  & & & & & & & & 162 & & 46 & 51 & 6 & 155 \\
   \cdashline{9-14}
&  & & & & & & & 140 & 21 & 42 & 65 & 7 & 160\\
  \hline \multirow{3}{*}{\#17} & \multirow{3}{*}{RS-10} & \multirow{3}{*}{20} & \multirow{3}{*}{1} & \multirow{3}{*}{[29, 46]} & \multirow{3}{*}{4} & \multirow{3}{*}{18} & \multirow{3}{*}{47} & 344 & \multirow{2}{*}{46} & 1227 & 97 & 10 & 5073 \\
  \cdashline{9-9}
 \cdashline{11-14}
  & & & & & & & & 391 & & 1329 & 82 & 7 & 4164 \\
   \cdashline{9-14}
&  & & & & & & & 289 & 42 & 1050 & 127 & 13 & 4770 \\
  \hline \multirow{3}{*}{\#18} & \multirow{3}{*}{V-Platoon III} & \multirow{3}{*}{30} & \multirow{3}{*}{-} & \multirow{3}{*}{[12, 21]} & \multirow{3}{*}{3} & \multirow{3}{*}{10} & \multirow{3}{*}{27} & 96 & \multirow{2}{*}{24} & 47 & 42 & 7 & 196 \\
  \cdashline{9-9}
 \cdashline{11-14}
  & & & & & & & & 107 & & 51 & 43 & 7 & 238 \\
   \cdashline{9-14}
  & & & & & & & & 91 & 22 & 48 & 52 & 8 & 240 \\ 
  \hline \multirow{3}{*}{\#19} & \multirow{3}{*}{V-Platoon III} & \multirow{3}{*}{30} & \multirow{3}{*}{1} & \multirow{3}{*}{[11, 23]} & \multirow{3}{*}{5} & \multirow{3}{*}{13} & \multirow{3}{*}{47} & 205 & \multirow{2}{*}{43} & 368 & 62 & 8 & 1298 \\
  \cdashline{9-9}
 \cdashline{11-14}
  & & & & & & & & 237 & & 393 & 67 & 7 & 2005 \\
   \cdashline{9-14}
  & & & & & & & & 220 & 42 & 381 & 92 & 11 & 2195 \\ 
   \hline \multirow{3}{*}{\#20} & \multirow{3}{*}{V-Platoon III} & \multirow{3}{*}{30} & \multirow{3}{*}{2} & \multirow{3}{*}{[11, 25]} & \multirow{3}{*}{7} & \multirow{3}{*}{15} & \multirow{3}{*}{65} & 373 & \multirow{3}{*}{63} & 1581 & 77 & 9 & 4855 \\
  \cdashline{9-9}
 \cdashline{11-14}
  & & & & & & & & 423 & & 1712 & 82 & 8 & 7046 \\
   \cdashline{9-9}
   \cdashline{11-14}
  & & & & & & & & 366 & & 1560 & 114 & 15 & 7893 \\  
  \hline
\end{tabular}
\caption{Evaluation results for complete characterization of counterexamples in linear systems with (without) bounded inputs. The bounded inputs value `-' denotes constant/no inputs. \textit{Dims} is the number of system variables; time bound $\timebound$ for Systems \#1 and \#2 is $15$ and for the rest of the systems is $100$. \textit{Verification Time} is the time \texttt{HyLAA} takes to compute the reachable set. $\mathrm{T_c}$ is the time taken by \algnm{\algfnt{construct\_bdd}} Algorithm. $C_{\unsafe}$ denotes the number of characterizations for a given safety violation in each system. Further, in each benchmark, the first row is for default ordering $O_d$, the second one is for $O_m$ ordering, and the third row corresponds to the random ordering $O_r$. The results highlighted in red are under-approximations of complete characterization incurred due to numerical in-stability.}
\label{table:complete-bdd}
\end{table*}


\textbf{Implementation:} The proposed algorithms have been implemented in a  linear hybrid systems verification tool, HyLAA. Simulations for reachable sets are performed using \texttt{scipy's odeint} function, which can handle stiff and non-stiff differential equations. Linear programming is performed using the GLPK library, and matrix operations are performed using \texttt{numpy}. We use Gurobi~\cite{gurobi} as the optimization solver for  MILP. The measurements are performed on a system running Ubuntu 16.04 with a 3.0 GHz Intel Xeon E3-1505M CPU with 8 cores and 32 GB RAM.

\textbf{Benchmarks:} We evaluate our techniques on various examples.  ``V-Platoon'' is a standard benchmark from  \textsc{Arch}
~\cite{benchmarksurl} suite; the ones annotated as ``RS'' are adopted from RealSyn~\cite{2018_realSyn}. We take the dynamics matrix $\mathcal{A}$ and the input matrix $\mathcal{B}$ from the original benchmark for each system. We modify the set of possible inputs, $\mathcal{U}$, and the safety specification suitable for our work in order to emphasize the complexity of the presented approach. The setup details are available at the github repository: \url{https://github.com/manishgcs/control-hylaa}. 

\textbf{Variables ordering:} We apply 3 different orderings of decision variables to construct 3 different decision diagrams for performance evaluation. For a monotonically increasing sequence of numbers $[1, 2, 3, 4, 5]$, the variables in default ordering $O_d$ are arranged as per the sequence. The second ordering $O_r$ is \emph{random} where elements are arranged in a random order. We make an observation about the reachable set overlap with the unsafe set for the third kind of ordering $O_m$. 
In most of the cases, we observe that the star which goes the deepest in the unsafe set lies in the middle of the sequence of the unsafe stars, thus it tends to have non-empty intersection with relatively more number of  stars in its neighborhood. We use this intuition to obtain another ordering  $[3, 2, 1, 4, 5]$ that has only the first half of the given sequence reversed which, in turn, brings the middle element to the front.

\textbf{Performance:} Table~\ref{table:complete-bdd} demonstrates our evaluation results. We measure the performance of our algorithms in terms of BDD creation time ($\mathrm{T_c}$), total number of nodes ($\mathcal{N}$), and its width ($\mathcal{W}$). The verification time for computing the reachable set is significantly less as compared to $\mathrm{T_c}$. Further, reduced BDD construction\footnote{We apply only  \emph{isomorphism} rule to our OBDD.} time takes much more time because we solve multiple optimization problems to check equivalence among nodes. The BDD construction time for systems with inputs is much higher due to addition of $m$ input variables at each successive step. Larger the time step at which the reachable set intersects with the unsafe set, higher is the number of continuous variables in the model  solved for feasibility or equivalence. The total number of continuous variables in the model for System \#17 is $66$ where $20$ is the number of state variables and  $46$ is the number of input variables due to the  largest time step at which the reachable set has a non-empty intersection with the unsafe set.

In most OBDD runs, the random ordering $O_r$ turns out to be slightly better than its counterparts in the BDD size. On the other hand, $O_m$ fares better in most ROBDD experiments, which supports our observation about the middle element likely having non-empty intersection with multiple elements. 

In Benchmark \#13 and \#20, the number of nodes ($\mathcal{N}$) are reduced by $75\%-80\%$ for $O_m$ across OBDD and ROBDD runs while the width ($\mathcal{W}$) is reduced by $80\%-85\%$. On the other hand, in benchmarks \#7 and \#18, we only see a reduction by $50\%$ in size for $O_m$ runs. If we take the summation of the total number of nodes of all benchmarks, we observe an overall reduction of $68 \%$ in $\mathcal{N}$ across OBDD and ROBDD runs. Similarly, we notice an overall reduction of $75\%$ in the width.

\begin{table}
\small
\centering
\caption{\small $\%$-variations ($\sigma$) in the BDD size ($\mathcal{N}$) due to different orderings. An appropriate order can yield up to $20\%$ reduction in the original BDD and $40\%$ reduction in the reduced BDD size.}
\label{tab:variation-in-size}
\begin{tabular}{|*{6}{cc|c||cc|c|}}
\hline
System & \multicolumn{2}{c||}{$\sigma$} & System & \multicolumn{2}{c|}{$\sigma$} \\
\cline{2-3}
\cline{5-6}
& OBDD & ROBDD & & OBDD & ROBDD\\
\hline
\hline
\#1 & 8.9 & 7.1 & \#11 & 8.9 & 37.1 \\
\hdashline
\#2 & 17.0 & 5.6 & \#12 & 25.0 & 30.7 \\
\hdashline
\#3 & 6.2 & 42.1 & \#13 & 11.4 & 46.4 \\
\hdashline
\#4 & 15.9 & 40.4 & \#14 & 12.1 & 28.4 \\
\hdashline
\#5 & 6.3 & 37.0 & \#15 & 14.3 & 38.8 \\
\hdashline
\#6 & 19.0 & 33.0 & \#16 & 13.6 & 21.6 \\
\hdashline
\#7 & 10.8 & 33.3 & \#17 & 26.1 & 35.4 \\
\hdashline
\#8 & 25.4 & 44.6 & \#18 & 15.0 & 19.2 \\
\hdashline
\#9 & 15.2 & 28.4 & \#19 & 13.6 & 32.6 \\
\hdashline
\#10 & 20.1 & 27.9 & \#20 & 13.5 & 32.5 \\
  \hline
\end{tabular}

\end{table}

\textbf{Variations in the size:} We also report the $\%$-variation $\sigma$ in the BDD size ($\mathcal{N}$) for each system in Table~\ref{tab:variation-in-size}. We define $\sigma \deq ((\mathcal{N}_{max} - \mathcal{N}_{min}) \div \mathcal{N}_{max}) \times 100$, where $\mathcal{N}_{max}$ (or $\mathcal{N}_{min}$) is the highest (or lowest) number of nodes among any of the orderings $O_d, O_m$ and $O_r$. The table underlines the importance of the decision variables ordering. It demonstrates that the BDD size can vary significantly across different orderings, and one can potentially achieve up to 50\% reduction in the reduced BDD size with appropriate ordering. 

\textbf{Numerical in-stability:} Recall that we model node isomorphism as an optimization problem encoded as an MILP. Although it is not observed in our evaluations, numerical in-stability can occur in an MILP-based approach even with a state-of-the-art commercial solver. 
The model requires the definition of $M$, which is larger than any value, and an arbitrarily small $\epsilon > 0$.  
An $\epsilon$ smaller than the solver's tolerance may translate into $y''$ variables becoming $0$ in system~\ref{eq:model1-encode-final-sol1}. This results into \algnm{\algfnt{isomorphs}} algorithm returning $\bot$ even when the predicates are equivalent. Consequently, ROBDD may end up having same size as OBDD. On the other hand, a higher $\epsilon$ value can sometimes lead to \algnm{\algfnt{isomorphs}} returning $\top$ even when the predicates are not equivalent. As a result, a few infeasible characterizations can get introduced (over-approximation) and/or some feasible ones may get dropped (under-approximation) from the diagram.
%
%

\textbf{Application to additional specifications:} We have maintained so far that the  diagrams~\ref{fig:characterization} or~\ref{fig:reduced-bdd} represent complete characterization of the violation of the safety specification $\phi \deq \mathrm{G}_{[0, 15]} \neg p$. We argue that the same diagrams also subsumes the characterization of the violation of some other safety specifications such as $\phi' \deq \mathrm{F}_{[0, 5]} p \rightarrow \mathrm{G}_{[6, 15]} \neg p$ and $\phi'' \deq \mathrm{G}_{[0, 10]} \neg p \rightarrow \mathrm{F}_{[11, 15]} \neg p$. The specification $\phi'$ states that if something bad happens within first $5$ time steps then nothing bad would happen between time interval $[6-15]$. On the other hand, $\phi''$ states that if nothing bad happens in first $10$ time steps then something good will eventually happen in the interval $[11-15]$. This observation is important for $2$ key reasons - (i) existing verification artifacts can be reused for safety verification w.r.t. such specifications, and (ii) it is not required to construct  individual decision diagram for each such specification.

\textbf{Extension to hybrid systems:} A linear hybrid system has multiple \emph{modes} (also called \emph{locations}) of operation  and discrete transitions as switching function between modes. These transitions labelled by \emph{guard} condition induce non-determinism
 which can possibly lead to multiple paths in the reachable set. Therefore, the reachable set in linear hybrid systems is denoted in the form of a tree (rooted at $\Theta$) called $ReachTree$. We can construct a bdd for each path in $ReachTree$ for generating respective characterization of counterexamples. The detailed explanation on $ReachTree$ computation and counterexample generation in linear hybrid systems is present in~\cite{DBLP:journals/automatica/GoyalD20}.

\section{Conclusion and Future Work} 
In this paper, we defined the notion of complete characterization of counterexamples in linear systems, and provided a sound and complete algorithm for its computation. We represented all modalities of a given safety violation using a binary decision diagram. Furthermore, we introduced a technique to perform isomorphism on its nodes to reduce the size of the diagram by significant proportion. The implementation of proposed techniques as a feature  of a model checking tool makes the verification artifacts more useful to system engineers. The future task is to extend the presented work to the unsafe set defined as a set of propositions. Another direction is to explore a more efficient graphical representation for such characterizations.


\noindent \emph{Acknowledgements:} 
This material is based upon work supported by the Air Force Office of Scientific Research under award number FA9550-19-1-0288 and National Science Foundation (NSF) under grant numbers CNS 1935724, 2038960. 
Any opinions, findings, and conclusions or recommendations expressed in this material are those of the author(s) and do not necessarily reflect the views of the United States Air Force or National Science Foundation.

\bibliographystyle{plain} 
\bibliography{tac}

\end{document}